\documentclass[journal abbreviation, manuscript]{copernicus}

\usepackage{amsthm}
\usepackage{amssymb}
\usepackage{mathrsfs}
\usepackage{draftwatermark} 
\SetWatermarkText{Pre-print version}
\SetWatermarkScale{0.5}

\newcommand{\dFaV}{\ensuremath \frac{\partial F_a}{\partial v}}
\newcommand{\dTaV}{\ensuremath \frac{\partial \tau_a}{\partial v}}
\newcommand{\dFaOmg}{\ensuremath \frac{\partial F_a}{\partial \omega}}
\newcommand{\dTaOmg}{\ensuremath \frac{\partial \tau_a}{\partial \omega}}
\newcommand{\dFaBeta}{\ensuremath \frac{\partial F_a}{\partial \beta}}
\newcommand{\dTaBeta}{\ensuremath \frac{\partial \tau_a}{\partial \beta}}
\newcommand{\dTaW}{\ensuremath \frac{\partial \tau_w}{\partial w}}

\begin{document}

\nolinenumbers

\title{Damping analysis of Floating Offshore Wind Turbine (FOWT): a new control strategy reducing the platform vibrations}


\Author[1]{Matteo}{Capaldo}
\Author[2]{Paul}{Mella}

\affil[1]{TotalEnergies OneTech, Palaiseau, France}
\affil[1,2]{Ecole Polytechnique, Institut Polytechnique de Paris, Palaiseau, France}




\correspondence{Matteo Capaldo (matteo.capaldo@totalenergies.com)}

\runningtitle{TEXT}

\runningauthor{TEXT}

\received{}
\pubdiscuss{} 
\revised{}
\accepted{}
\published{}


\firstpage{1}

\maketitle

\begin{abstract}
In this paper, the coupled dynamics of the floating platform and the WTG rotor is analysed. 
In particular, the damping is explicitly derived from the coupled equations of rotor and floating platform. 
The analysis of the damping leads to the study of the instability phenomena and it derives the explicit conditions that lead to the Non Minimum Phase Zero (NMPZ). 
Two NMPZs, one related to the rotor dynamics and the other one to the platform pitch dynamics, are analysed. The latter is a novelty and it is analysed in this work, providing the community of an explicit condition for its verification. 
The domain of the instability of the platform is explicitly derived from the coupled system of equations. 
In the second part of the paper, from the analysis of the damping of the floating platform, a new strategy for the control of FOWTs is proposed. 
This strategy allows one to impose to the controller an explicit level of damping in the platform pitch motion without changing the period of platform pitching. 
Finally the new strategy is compared to the one without compensation by performing aero-hydro-servo-elastic numerical simulations of the UMaine IEA15MW FOWT. 
Generated power, movements, blade pitch and tower base fatigue are compared showing that the new control strategy can reduce fatigue in the structure without affecting the power production. 
\end{abstract}


\introduction  
Wind energy is an important source of renewable energy and it has a very high potential in both onshore and offshore. In terms of installed capacity, onshore wind is still the largest contribution. However, in the next years, the new annual offshore installed capacity is estimated to exceed 30 GW by 2030, in order to stay on-track for a netzero/$1.5^{\circ} \ C$ pathway \cite{gwec}. 
In offshore, there is a growing interest in floating foundations. In fact, FOWTs would allow access to good wind resource locations that are not
suitable for fixed-bottom foundations.

In that context, the levelized cost of energy (LCOE) of offshore wind farms needs to be decreased to be competitive with respect to onshore wind. This is especially true for the  FOWT. 
One effective way to achieve this objective is to investigate different strategies for the control of the FOWTs in order to improve it with respect to the LCOE. 
As explained in \cite{book_control}, the minimisation of the LCOE involves a series of partial objectives, energy capture, mechanical loads and power quality.
These objectives are actually closely related and sometimes conflicting and they should not be pursued separately. 
Hence, the question is to find a well balanced compromise among them.  
Considering FOWT, this optimization problem increases in complexity since the movements of the floating platform interact with the feedback control loop. 
Moreover, the coupling between the platform movements, the rotor dynamics and the blade pitch control can lead to oscillating (or not damped) stability or even to unstable conditions \cite{Larsen}. 

The nature and the set of control parameters leading to the instability can variate from one platform design to another one, e.g., barge, spare, semi-sub or tension-leg platform. 
However, for each of the platforms, there are sets of design and control parameters leading to oscillating stability or instability \cite{Fleming}. 
The issue come from the fact that in above rated wind speed, the blade pitch regulates the speed by increasing the angle of attack to feather. For a FOWT, when the platform has a forward motion, the rotor experiences an increasing wind speed. 
Consequently, the blade pitch control increases the angle of attack to feather and, hence, reduces the aerodynamic torque and regulates the rotor speed. 
However, it also reduce the rotor thrust that induces a further forward motion. 
So the blade pitch control amplifies the original forward motion of the platform because the floating platform surge and pitch natural frequencies  are in the bandwidth of the blade pitch control.

Solutions exist to avoid this phenomenon. 
The first one and the most common in literature is to reduce the blade pitch control proportional and integral gains in order to reduce its bandwidth and exclude the platform pitch and surge natural frequencies \cite{jon_pitch_gain, Larsen}.
However, this solution do not completely solve the problem and, moreover, the price to pay is to have a less reactive blade pitch control that allows important over speed of the rotor. 
Alternative methods use additional sensing, such as nacelle fore-aft acceleration measurements or platform gyroscopes, to improve the performance of the pitch controller.
In \cite{lack_platf_pitch} and \cite{OMAE2020-18770}, the platform pitch velocity is used to adjust the rated speed set point to reduce platform motions. 
However, the platform pitch damping analysis is not investigated and the link with the compensation parameter is not given. 
Higher order controllers, such as a linear quadratic regulator (LQR) are applied and evaluated in \cite{LQR_Ma}. Disturbance accommodating controller (DAC) is evaluated here \cite{DAC} and it is coupled with individual pitch control (IPC) in this paper, \cite{Namik_IPC}. 
A nonlinear pitch and torque controller using wind preview is designed in \cite{nonlinear_MPC} and \cite{mpc_mata}, giving promising results.

The control strategy proposed in this work shares some similarities to the one introduced in \cite{lack_platf_pitch} and \cite{OMAE2020-18770}. 
As introduced in this paper, in fact, the rated generator speed is no longer a constant value but a function of the platform pitch velocity and the blade pitch is used to damp the floating platform pitch. 
However, differently from \cite{lack_platf_pitch}, the ratio between proportional and integral gains of the correction can be considered different for the platform pitch motion and the rotor speed. 
This choice, in fact, can be not optimal for some cases. 
Differently from \cite{OMAE2020-18770}, a mathematical frame is given to define the compensation to the platform pitch and the explicit form of it is analytically given.  
Moreover, a second compensation is considered in the control strategy proposed in this work. 
It is to solve the Non Minimum Phase Zero (NMPZ) effects caused by the blade pitch on the rotor rotational dynamics. 
It is already introduced in \cite{Stockhouse} and it is, in this work, analytically developed. 
The study of the NMPZs brings to a new NMPZ phenomenon, described for the first time in this work. 
This is the NMPZ caused by the blade pitch on the platform dynamics.
This new phenomenon is analytically developed leading to the explicit condition to verify it. However, the compensations proposed by this model can't correct it. 

The novelty of this work is related to the FOWT damping analysis, i.e., the damping obtained by coupling the rotor and the platform pitch dynamics. 
This damping is explicitly derived from the coupled equations of rotor and floating platform. 
This analysis leads to the study of the instability phenomena underlining the conditions leading to the NMPZ. 
One new NMPZ, never discussed in literature, is discovered and analysed in this work. 
The domain of the instability of the platform is explicitly derived from the coupled system of equations. 
The control strategy proposed relies on this analysis and it allows to impose an explicit level of damping in the platform pitch motion to the controller without changing the period of platform pitching. 
This explicit form of the damping in the platform pitch dynamics is a novelty of this work. 

The chosen strategy is, then, compared to the one which does not consider a platform pitch compensation. 
The latter neglects the following phenomenon: the blade feather reduces the aerodynamic thrust and, thus, a part of the opposing force on the platform is neglected.

The document is organized as follows. In Section \ref{sec:fowt_equations}, the equations of the FOWT system are described with the considered degrees of freedom and their coupling terms. 
Section \ref{sec:control_model} presents the control model considered in this work.
The closed loop feedback system is then analysed in Section \ref{NMPZ}, 
leading to the definition of the boundaries of the instability domain.
For this controller, a new control strategy dedicated to FOWTs, 
named $ \zeta_{plt} $-fixed is presented and analytically derived in Section \ref{sec:damping} and \ref{Ptfm}. 
Some numerical tests are presented and commented in Section \ref{numerical}.

\section{Floating Offshore Wind Turbine and its controller}

\label{sec:fowt_equations}
The floating offshore wind turbine (FOWT) is represented by a system of two degrees of freedom, namely the rotor speed $\Omega$ and the platform pitch angle, $\Phi$. Two control parameters, $B$ (blade pitch) and $\tau_g$ (generator torque), and two external disturbances, $V$ (wind speed) and $W$ (waves speed) are considered. 
For all the values that form a given operating point (namely $\Omega, \Phi, B, T_g, V, W$), the notation $X = \Bar{x} + x$ is adopted, with $x$ being the small perturbation of a steady-state operating point $\Bar{x}$. 
\\

The model is, then, based on the two fundamental equations:
\begin{equation}
    \label{eq1} \dot \Omega_g = \frac{N_g}{J_r} (T_a - N_g T_g) 
\end{equation}
\begin{equation}
    \label{eq2} J_t \ddot \Phi + D_t \dot \Phi + K_t \Phi = h_t F_a + \tau_w
\end{equation}
where $\Omega_g$ is the generator speed, hereafter noted $\Omega$, $T_a$ and $T_g$ are the aerodynamic and electric torque, $N_g$ is the gearbox ratio and $J_r$ is the rotor inertia. 
$J_t$ is the total system moment of inertia about the pitch rotation, $D_t$ is the natural damping coefficient (assumed constant), $K_t$ is a spring-like restoring coefficient (mainly given by the mooring lines of the floating platform),
$h_t$ is the height of the rotor (approximately the tower length), $F_a$ is the aerodynamic force flowing from the rotor to the system and $\tau_w$ is the overturning moment given by the waves. 
\\
Once a steady-state operating point $\Bar{x}$ is reached, the same two equations can be applied to any small variations $x$ around this operating point. Equations \ref{eq1} and \ref{eq2} applied on $X$ can be written:
$$ \dot \omega = \frac{N_g}{J_r} (\tau_a - N_g \tau_g) $$
$$ J_t \ddot \phi + D_t \dot \phi + K_t \phi = h_t dF_a + d\tau_w$$
The infinitesimal thrust and torques satisfy $T_g = \Bar{\tau}_g + \tau_g$, $T_a = \Bar{\tau}_a + \tau_a$, $F_a = \Bar{F_a} + dF_a$ and $\tau_w = \Bar{\tau}_w + d \tau_w$. 
By considering small perturbations of a steady-state operating point (given by $\Omega, \Phi, B, T_g, V, W$), it allows one to use the following linear forms:
$$ \tau_a = \dTaOmg \omega + 
            \dTaV v_r + 
            \dTaBeta \beta$$
$$d F_a = \dFaOmg \omega + 
            \dFaV v_r + 
            \dFaBeta \beta$$
$$d \tau_w = \dTaW w$$

Note that $F_a$ and $\tau_w$ are separated for clarity, which explains why $dF_a$ does not depend on a $w$ perturbation. Moreover, the hypothesis of $\phi$ being small allows one to remove the terms $\frac{\partial \tau_a}{\partial \phi} \phi$, $\frac{\partial F_a}{\partial \phi} \phi$ and $\frac{\partial \tau_w}{\partial \phi} \phi$.

Relative wind $v_r$ is the wind velocity in rotor reference frame, it is computed from $v$ by:
$$v_r = v - h_t \dot \phi$$
Under the assumption of $\phi$ being small, $h_t \dot \phi$ represents the rotor fore-aft velocity in a fixed global reference frame.

Equations \ref{eq1} and \ref{eq2} applied to small perturbations of a steady-state point can therefore be expressed in the linear form:
\begin{equation}
\label{eq1_p}
    \tag{1'} \dot \omega = \frac{N_g}{J_r} \left(\dTaOmg \omega + 
            \dTaV (v - h_t \dot \phi) + 
            \dTaBeta \beta - N_g \tau_g \right) 
\end{equation}
\begin{equation}
\label{eq2_p}
    \tag{2'} J_t \ddot \phi + D_t \dot \phi + K_t \phi = h_t \left(\dFaOmg \omega + 
            \dFaV (v - h_t \dot \phi) + 
            \dFaBeta \beta \right) + \dTaW w
\end{equation}
Those coupled second order equations yield the following four dimensional state-space model: 
\begin{equation}
    \label{eq3} \dot x = A_0 x + B_c u_c + B_d u_d
\end{equation}
Where $x = (\theta, \dot \theta, \phi, \dot \phi)^T$, $\theta = \int \omega$ (i.e. $\dot \theta = \omega$), $u_c = (\beta, \tau_g)^T$ and $u_d = (v, w)^T$, and : 
\\
$$
A_0 =   \begin{bmatrix}
        0 & 1                       & 0                 & 0                                 \\
        0 & \frac{N_g}{J_r} \dTaOmg & 0                 & -h_t \frac{N_g}{J_r} \dTaV        \\
        0 & 0                       & 0                 & 1                                 \\
        0 & \frac{h_t}{J_t} \dFaOmg & - \frac{K_t}{J_t} & -\frac{1}{J_t} (D_t + h_t^2 \dFaV)\\
        \end{bmatrix}
\qquad
B_c =   \begin{bmatrix}
        0                       & 0                         \\
        \frac{N_g}{J_r} \dTaBeta   & -\frac{N_g^2}{J_r} \\
        0                       & 0                         \\
        \frac{h_t}{J_t} \dFaBeta   & 0  \\
        \end{bmatrix}
B_d =   \begin{bmatrix}
        & 0                 & 0                 \\
        &  \frac{N_g}{J_r} \dTaV  & 0                 \\
        & 0                 & 0                 \\
        & \frac{h_t}{J_t} \dFaV             & \frac{1}{J_t} \dTaW  \\
        \end{bmatrix}
$$

\subsection{Control model description}
\label{sec:control_model}
In this section the pitch controller model is described. 

The present control model considers $\omega_r$ as the reference for the $\omega$ and $0$ as the reference for $\dot \phi$. It is based on several SISO (single-input-single-output) feedback loops. It can be seen as a multi-SISO:

\begin{itemize}
\item Proportional : $ \beta_{P} = k_P (\Omega - \Omega_r)$  
\item Integral :  $ \beta_{I} = k_I \int (\Omega - \Omega_r) = k_I (\theta - t \Omega_r)$
\item Beta platform pitch compensation : $ \beta_{comp} = k_{\beta} ( \dot \Phi - \dot \Phi_r )$
\item Generator torque platform pitch compensation : $ \tau_{g, comp} = k_{\tau_g} ( \dot \Phi - \dot \Phi_r )$
\end{itemize}

Controllers described by the literature considering the same compensations \cite{rosco,Stockhouse} aim at maintaining $\omega$ steady near its rated value
by acting on the blade pitch $\beta$ to variate the aerodynamic torque $\tau_a$ with the opposite sign with respect to the rotor relative speed $\omega$. 
However, this strategy neglects the following phenomenon: the blade feather modifies the aerodynamic thrust $F_a$. Thus, a part of the opposing force on the platform is neglected. 
The strategy developed in this paper aims at minimizing $\phi$ variations with the constraint of maintaining a constant $\omega$.
Such a control strategy should reduce the loads on the structures (nacelle, tower and floater). 
Section \ref{numerical} considers a full aero-hydro-servo-elastic model to verify this assumption. It is analysed the performance of the control strategy in a FOWT realistic environment reproduced by a numerical twin.

\subsection{Global State-Space description}

\label{SS description}

For a FOWT, the objective ot the pitch control is to remain in the equilibrium operating point.
It translates in:  $\bar \omega = \Omega_r$ and $\bar {\dot {\phi}} = \dot \Phi_r = 0$. 
This objective allows one to justify the linear form for of the global equations \ref{eq1_p} and \ref{eq2_p}.
For constant inputs $\bar v$ and $\bar w$, this operating point is reached by the appropriated constants $\bar \beta$ and $\bar{\tau_g}$, the fine pitch and fine electric torque, respectively.

The control model is, here, introduced into the wind turbine state space description.
For small perturbations of this steady-state operating point, the PI controller described previously becomes :

\begin{itemize}
\item Proportional: 
\begin{equation}
\label{eq:cont_mod_p}
\beta_{P} = k_P \omega
\end{equation}
\item Integral:  
\begin{equation}
\label{eq:cont_mod_i}
\beta_{I} = k_I \int \omega = k_I \theta
\end{equation}
\item Beta platform pitch compensation:
\begin{equation}
\label{eq:cont_mod_compb}
\beta_{comp} = k_{Beta}  \dot \phi
\end{equation}
\item Generator torque platform pitch compensation: 
\begin{equation}
\label{eq:cont_mod_compt}
\tau_{g, comp} = k_{\tau_g}  \dot \phi 
\end{equation}
\end{itemize}

\begin{figure}[H]
\centering
\includegraphics[scale=0.33]{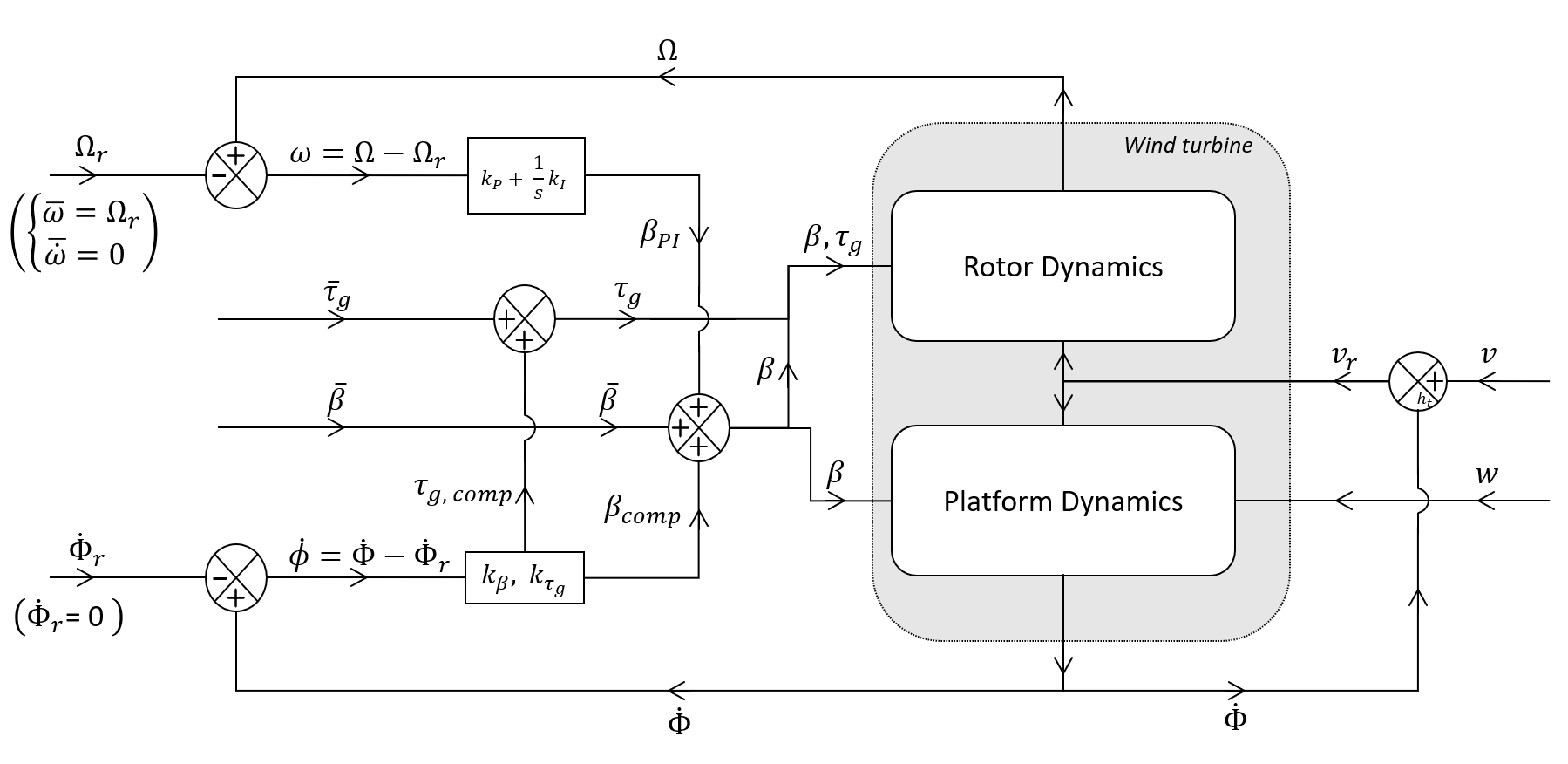}
\caption{Block diagram of the controller model}
\label{fig:figure_diagram}
\end{figure}

Figure \ref{fig:figure_diagram} shows the entire picture of the control model. 
This control strategy acts on the two dynamic systems, platform and rotor. 
Hence, one can appreciate how the bottom-fixed scheme acting on the rotor ($\omega$) with a proportional integral scheme is then corrected by the $\beta_{comp}$ and the $\tau_g$ that depends on the platform pitch dynamics error.

The linear expression of $u_c = (\beta, \tau_g)^T$ as a function of $x = (\theta, \dot \theta, \phi, \dot \phi)^T$ is $u_c = K_0 x + u_{c, ol}$ where:
$$
K_0 =   \begin{bmatrix}
        k_I & k_P & 0 & k_{\beta}   \\
        0   & 0   & 0 & k_{\tau_g}  \\
        \end{bmatrix}
$$
is the matrix of the control gains and $u_{c, ol}$ is an optional additional control (open loop) that can be considered. This is useful to analyse the NMPZ in the next section. 
By replacing it in equation \ref{eq3}, it leads to:
\begin{equation}
    \label{eq4} 
    \dot x = (A_0 + B_c K_0) x + B_c u_{c, ol} + B_d u_d
\end{equation}
Which leads to define the global matrix of the system of equations: 
\begin{equation}
\label{eq4Matrix} 
A = (A_0 + B_c K_0) = \begin{bmatrix}
        0                            & 1                                        & 0                 & 0                                 \\
        k_I \frac{N_g}{J_r} \dTaBeta & \frac{N_g}{J_r} (\dTaOmg + k_P \dTaBeta) & 0                 & \frac{N_g}{J_r} (-h_t \dTaV + k_\beta \dTaBeta - k_{\tau_g} N_g)\\
        0                            & 0                                        & 0                 & 1                                 \\
        k_I \frac{h_t}{J_t} \dFaBeta & \frac{h_t}{J_t} (\dFaOmg + k_P \dFaBeta) & - \frac{K_t}{J_t} & \frac{1}{J_t} (-D_t - h_t^2 \dFaV + k_\beta h_t \dFaBeta)\\
        \end{bmatrix}
\end{equation}

The time domain system can be rewritten in the complex domain. Using the following notation, $\mathscr{L}\{x(t) \} = \underline{X}$,
equation \ref{eq4} translates into:

\begin{equation}
\underline{X} = (sI - A)^{-1} (B_c \underline{U}_{c, ol} + B_d \underline{U}_d)
\end{equation}
This leads us to define 
$$B = \begin{bmatrix}
    B_c & | & B_d
    \end{bmatrix}
, 
\;
u = \begin{pmatrix}
    \beta_{ol} \\
    \tau_{g, ol} \\
    v \\
    w \\
    \end{pmatrix}
\;
$$
and 
\begin{equation}
G(s) = (sI - A)^{-1} B = \frac{1}{det(sI - A)} Com(sI - A)^T B = \frac{1}{\chi_A(s)} N(s)
\end{equation}
in order that :
\begin{equation}
    \label{eq5} 
    \chi_A(s) \underline{X} (s) = N(s) \underline{U} (s)
\end{equation}

G(s) is a $4 \times 4$ matrix of which every component can be written as the quotient of a polynomial in s and $\chi_a(s)$.

\subsection{Non minimum phase zeros analysis and resolution (negative damping on the control)}

\label{NMPZ}

In this section, we address the problem of negative damping by addressing the positions of the zeros of each component of G in the complex plane, i.e. the points where equation \ref{eq5} is well defined but becomes: 
\begin{equation}
\chi_A \ X_0 ^T \cdot  \underline{X} = 0. 
\end{equation}

This translates the fact that $s$ is a NMPZ if it exists a specific $X_0 ^T$, such as, for any value of $\underline{U}(s)$, the linear combination $X_0 ^T N(s) \underline{U}(s)$ gives $X_0 ^T = 0$.

Physically a $X_0 ^T$ is equivalent to an infinitesimal shifting along a specific direction of a steady-state point that can not be obtained with any infinitesimal shifting of the input. This phenomenon is better illustrated case by case.
\\
For the rest of the section, an open loop control on $\beta$, is considered in order to highlight the NPMZ. 
Hence, $\beta_{ol}$ is added to the multiple SISO as already described in \ref{eq4}.
Since the feedback control on $\beta$ can't erase the NMPZ condition, to lighten the formulas, the notation $\beta$ will be used instead of $\beta_{ol}$ in this section. 

\paragraph{$\phi$-NMPZ: negative damping on $\phi$ control by $\beta$}

Gain equation in the Laplace domain for $\beta \rightarrow \phi$ control is obtained by projecting \ref{eq5} on the $x = (0, 0, \phi, 0)$ axis and considering only a $\beta$ perturbation, ie. an input $u = (\beta_{ol}, 0, 0, 0)$. The resulting equation is:

\begin{equation}
    \label{eq6} 
  \chi_A(s) \underline \phi(s) = N_{3, 1}(s) \underline \beta (s),  
\end{equation}

\begin{equation}
\label{eq6b} 
 N_{3, 1}(s) = \frac{J_r}{N_g} \dFaBeta s^2 + \left(\dTaBeta \dFaOmg - \dFaBeta \dTaOmg \right) s 
\end{equation}

The condition for the NMPZ on $\beta \rightarrow \phi$ control is that $N_{3,1}$ has a root with a real part strictly positive. 
Assuming that $\beta = \beta_f $, with $\beta_f$ the fine pitch, the previous derivatives are all negative.
Hence, the root research of $N_{3,1}$ leads to:

\begin{equation}
\label{eq:cond_nmpz}
 \dTaOmg \ / \  \dTaBeta \quad < \quad \dFaOmg \  /  \  \dFaBeta
\end{equation}

Intuitively, this corresponds to an operating point where \textit{$\tau_a$ is rather influenced by $\beta$} and \textit{$F_a$ is rather influenced by $\omega$}. 
This NMPZ does not depend on parameters of the platform, it is only related to WTG performances. 
However, the amplitude of the phenomenon is related to the platform properties. 
It is to be noted that this NMPZ has never been highlighted in literature and the control model (with compensations of the platform motions) introduced in Section \ref{sec:control_model} in the feedback control loop does not prevent it. 
It results only from the characteristic of the FOWT system. 
Further works should focus on this phenomenon and introduce corrections to prevent it for any FOWT system. 

In absence of NMPZ, ie. eq. \ref{eq:cond_nmpz} being false, increasing $\beta$ from a steady-state operating point (ie. setting $d \ddot \beta > 0$, $d \dot \beta > 0$ and $d \beta > 0$) will always imply a reduction of $\phi$. 
In presence of NMPZ, ie. if eq. \ref{eq:cond_nmpz} is true, the reduction or the increase of $\phi$ depends on the ratio between $\ddot \beta$ and  $\dot \beta$.

\begin{figure}[H]
\centering
\includegraphics[scale=0.6]{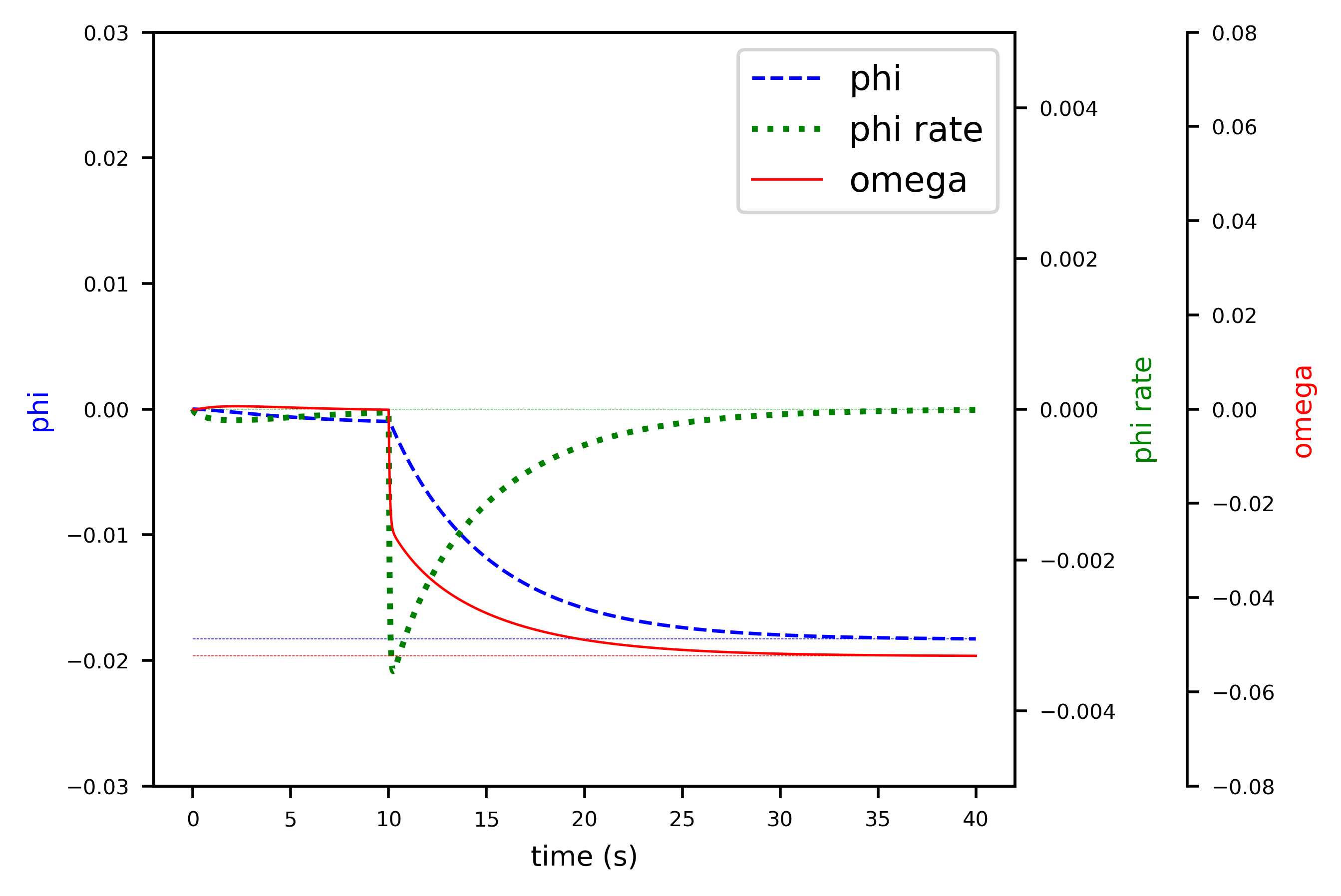}
\includegraphics[scale=0.6]{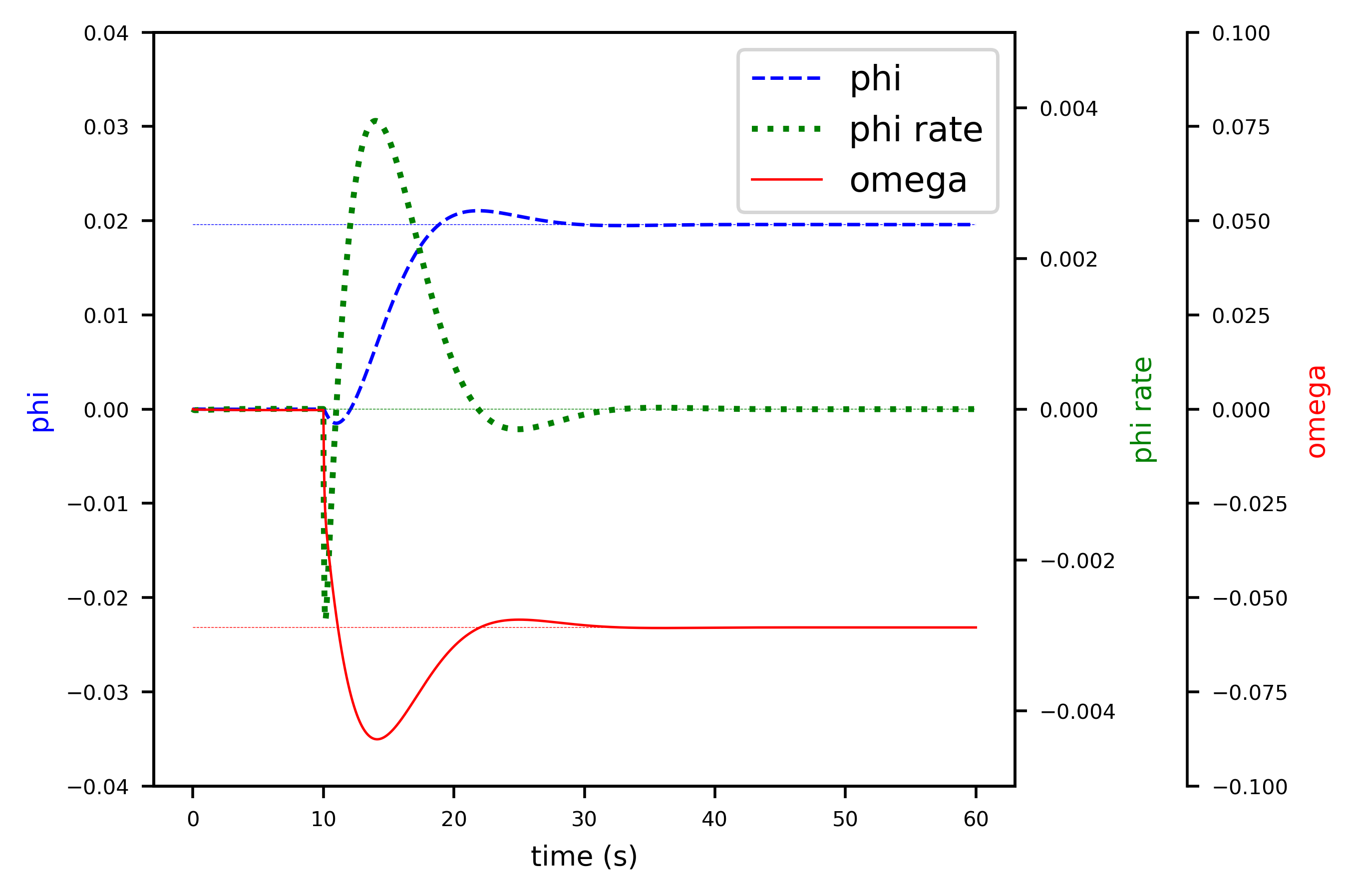}
\caption{$\phi$ and $\omega$ responses to a $\beta$-step input (at $t = 10 s$): on the left, values (Table \ref{tab:nmpz1}) are chosen so that eq. \ref{eq:cond_nmpz} is false: $\phi$ decreases. 
On the right, values (Table \ref{tab:nmpz1}) are chosen so that the eq. \ref{eq:cond_nmpz} is true: $\phi$ increases even though $\beta$ has step up.}
\label{fig:figure_nmpz1}
\end{figure}

The latter only happens when eq. \ref{eq:cond_nmpz} is verified: increasing blade pitch  \textit{reduces $\tau_a$ more than it increases $F_a$} (because  \textit{$\tau_a$ is rather influenced by $\beta$}), 
thus $\omega$  increases and causes $F_a$ to decrease (because \textit{$F_a$ is rather influenced by $\omega$}) and then $\phi$ to increase. 
If eq. \ref{eq:cond_nmpz} is not verified, increasing blade pitch $\beta$ from a steady-state operating point always reduces platform pitch $\phi$.
In practice, this effect can become an issue for a control algorithm mainly focused on $\omega$ stabilization since it generates unexpected platform dynamics. 
Figure \ref{fig:figure_nmpz1} reproduces in the time domain $\phi$ and $\omega$ responses to a $\beta$-step input (at $t = 10 s$):  values (resumed in Table \ref{tab:nmpz1}) are chosen so that eq. \ref{eq:cond_nmpz} is false: $\phi$ decreases. 
On the right, values  are chosen so that the eq. \ref{eq:cond_nmpz} is true: $\phi$ increases even though $\beta$ has step up.

\begin{table}[ht]
\centering
\caption{Set of parameters to show the NMPZ of eq. \ref{eq:cond_nmpz} }
\begin{tabular}[t]{ccc}
 \hline  
 $  $ & eq. \ref{eq:cond_nmpz} false & eq. \ref{eq:cond_nmpz} true \\ 
 \hline
 $\dTaV$ & $2980.9 ~ kN.s$ &   $3079 ~ kN.s $\\ 
 \hline
 $\dFaV$ & $354.8 ~ kN.s.m^{-1}$  &  $355.6 ~ kN.s.m^{-1} $ \\ 
 \hline
 $\dTaOmg$ & $-58597.1 ~ kN.m.s.rad^{-1}$ &   $ -55499.5 ~ kN.m.s.rad^{-1} $\\ 
 \hline
 $\dFaOmg$ & $-5658.0 ~ kN.s.rad^{-1}$ & $-5820.4 ~ kN.s.rad^{-1} $ \\
 \hline
 $\dTaBeta$ & $-152347.8 ~ kN.m.rad$ &  $ -160140.5 ~ kN.m.rad$ \\
 \hline
 $\dFaBeta$ & $-16052.2 ~ kN.rad$ &  $ -15260 ~ kN.rad$ \\
\hline
\end{tabular}
\label{tab:nmpz1}
\end{table}%

\paragraph{$\omega$-NMPZ: negative damping on $\omega$ control by $\beta$}

Gain equation in the Laplace domain for $\beta \rightarrow \omega$ control is given by:
\begin{equation}
\chi_A(s) \ \underline \omega (s) = N_{2, 1}(s) \ \underline \beta (s)  
\end{equation}
\begin{equation}
N_{2, 1}(s) = \frac{J_t}{h_t} \dTaBeta s^3 + \left[\frac{D_t}{h_t} \dTaBeta + h_t \left(\dTaBeta \dFaV - \dFaBeta \dTaV \right) - k_{\tau_g} N_g \dFaBeta \right] s^2 + \frac{K_t}{h_t} \dTaBeta s
\end{equation}

Hence the condition for NMPZ on $\beta \rightarrow \omega$ control is:

\begin{equation}
    \label{eq:cond_nmpz2}
    h_t^2 \left(\dFaV - \left(\dTaV + k_{\tau_g} \frac{N_g}{h_t} \right)  \frac{\dFaBeta}{\dTaBeta} \right) < - D_t
\end{equation}

This corresponds with an operating point where \textit{$\tau_a$ is rather influenced by $v$} and \textit{$F_a$ is rather influenced by $\beta$}. 
In presence of NMPZ, i.e. if eq. \ref{eq:cond_nmpz2} is true, the sign of $d\omega$ depends on the choice of $d \ddot \beta$, $d \dot \beta$ and $d \beta$.
Intuitively, the latter only happens when eq. \ref{eq:cond_nmpz2} is verified: 
increasing blade pitch will \textit{reduce $F_a$ more than it increases $\tau_a$} (because  \textit{$F_a$ is rather influenced by $\beta$}), 
thus $\dot \phi$ will decrease and cause relative wind $v_r = v - h_t \dot \phi$ to increase. As \textit{$\tau_a$ is rather influenced by $v$}, this will reduce $\omega$ in the end.
In practice, this effect can become an issue if a $\omega$ control algorithm obtains the opposite result than what was expected.

Because of this issue, literature has concerned the NMPZ phenomenon on $\beta \rightarrow \omega$ control and suggested several control corrections. 
Due to the nature of this phenomenon, any correction concerning $\beta$ control, introduced in eq. \ref{eq:cont_mod_compb}, can't prevent this NMPZ. 
However, detuning the PI controller (by lowering $k_P$ and $k_I$ gains) or using the $\beta$ platform pitch compensation $d\beta_{comp} = k_\beta \dot \phi$ as suggested in  \cite{rosco}, can mitigate the effect of NMPZ when eq. \ref{eq:cond_nmpz2} is true. 

However, the complete prevention of the problem can be obtained by several set of parameters that involves both the WTG, 
the floating platform and the control set-up. 
In fact, for this NMPZ, equation \ref{eq:cont_mod_compt} of the control model described in Section \ref{sec:control_model} allows one to avoid NMPZ by choosing a well-suited value of  $k_{\tau_g}$. 
This compensation has been already introduced by \cite{Stockhouse} with the formula:
\begin{equation}
    \label{ktau_g}
    k_{\tau_g} = - m_{\tau_g} \frac{h_t}{N_g} \dTaV, ~ ~ ~ m_{\tau_g} \in [0, 1]
\end{equation}
The authors notice however it usually needs to be saturated because of turbine generator design constraints. 
In absence of NMPZ, i.e. eq. \ref{eq:cond_nmpz2} being false, increasing $\beta$ from a steady-state operating point will always imply reducing $\omega$.

\begin{figure}[H]
\centering
\includegraphics[scale=0.6]{./plots/NMPZ/NMPZneither}
\includegraphics[scale=0.6]{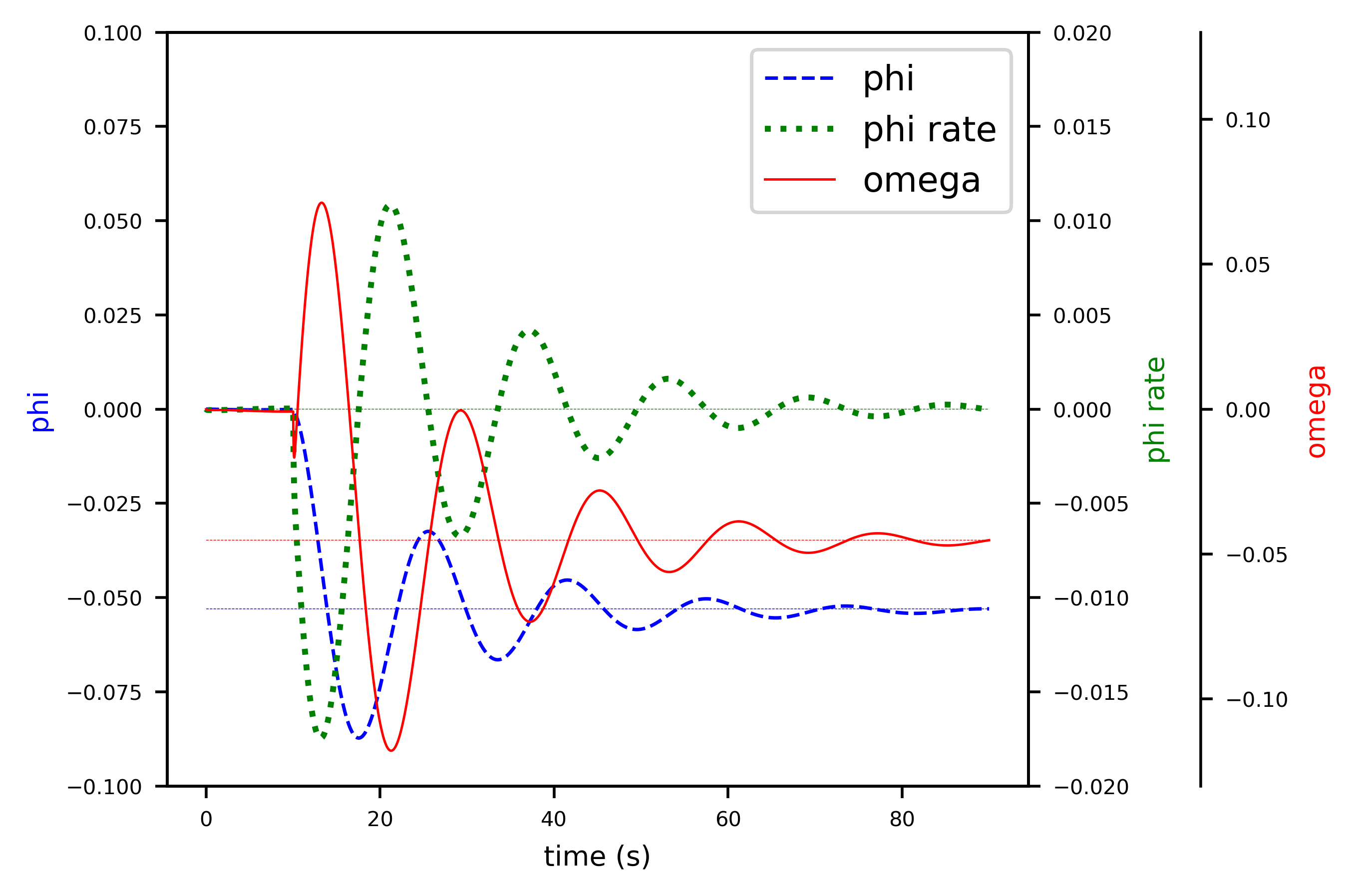}
\caption{$\omega$ responses to a $\beta$-step input (at $t = 10 s$). 
On the left, parameters (Table \ref{tab:nmpz2}) are chosen so that the condition eq. \ref{eq:cond_nmpz2} is false: $\omega$ decreases. 
On the right, parameters are chosen so that condition eq. \ref{eq:cond_nmpz2} is true: at first $\omega$ increases even though $\beta$ has step up.}
\label{fig:NPMZ2}
\end{figure}

In order to visualize this NMPZ, Figure \ref{fig:NPMZ2} shows $\omega$ responses to a $\beta$-step input (at $t = 10 s$). 
On the left, parameters (Table \ref{tab:nmpz2}) are chosen so that the condition eq. \ref{eq:cond_nmpz2} is false: $\omega$ decreases. 
On the right, parameters are chosen so that condition eq. \ref{eq:cond_nmpz2} is true: at first $\omega$ increases even though $\beta$ has step up.

\begin{table}[ht]
\centering
\caption{Set of parameters to show the NMPZ of eq. \ref{eq:cond_nmpz2} }
\begin{tabular}[t]{ccc}
 \hline  
 $  $ & eq. \ref{eq:cond_nmpz2} false & eq. \ref{eq:cond_nmpz2} true \\ 
 \hline
 $\dTaV$ & $2980.9 ~ kN.s$ &   $ 2838 ~ kN.s $\\ 
 \hline
 $\dFaV$ & $354.8 ~ kN.s.m^{-1}$  &  $303.0 ~ kN.s.m^{-1} $ \\ 
 \hline
 $\dTaOmg$ & $-58597.1 ~ kN.m.s.rad^{-1}$ &   $ -59428.7 ~ kN.m.s.rad^{-1} $\\ 
 \hline
 $\dFaOmg$ & $-5658.0 ~ kN.s.rad^{-1}$ & $ -6282.9 ~ kN.s.rad^{-1} $ \\
 \hline
 $\dTaBeta$ & $-152347.8 ~ kN.m.rad$ &  $ -133058.7 ~ kN.m.rad$ \\
 \hline
 $\dFaBeta$ & $-16052.2 ~ kN.rad$ &  $ -18247.0 ~ kN.rad$ \\
\hline
\end{tabular}
\label{tab:nmpz2}
\end{table}%

\paragraph{Conclusion of NMPZ analysis:}

Comparison between Figures \ref{fig:NPMZ} and \ref{fig:NPMZ2} enlightens what really happens after a step input, with and without NMPZ: at the beginning both $\omega$ and $\dot \phi$ always decrease just after the step. 
However, when both NMPZ conditions eq. \ref{eq:cond_nmpz} and eq. \ref{eq:cond_nmpz2} are false, those tendencies don't change. 
Conversely, when eq. \ref{eq:cond_nmpz} is true, we observe that $| \dot \omega |$ is so big that $\dot \phi$ jumps into positive values. Similarly, when  eq. \ref{eq:cond_nmpz2} is true, we observe that $| \dot \phi |$ is so big that $\omega$ jumps (only for a short time) into positive values. NMPZ, as we have seen in the examples, can cause important shifts and unexpected behaviors for both $\omega$ and $\phi$. 

The NMPZ $\beta \rightarrow \phi$ doesn't depend on above defined parameters: 
the model in Section \ref{sec:control_model} does not give solution to always prevent it,
but condition eq. \ref{eq:cond_nmpz} forecasts which operating points it affects. 
On the other hand, a wise choice of $\tau_g$ avoids $\beta \rightarrow \omega$ NMPZ, 
which is the main reason why this compensation has already been introduced by \cite{Stockhouse} and in the control model in Section \ref{sec:control_model}.

\begin{figure}[H]
\centering
\includegraphics[scale=0.65]{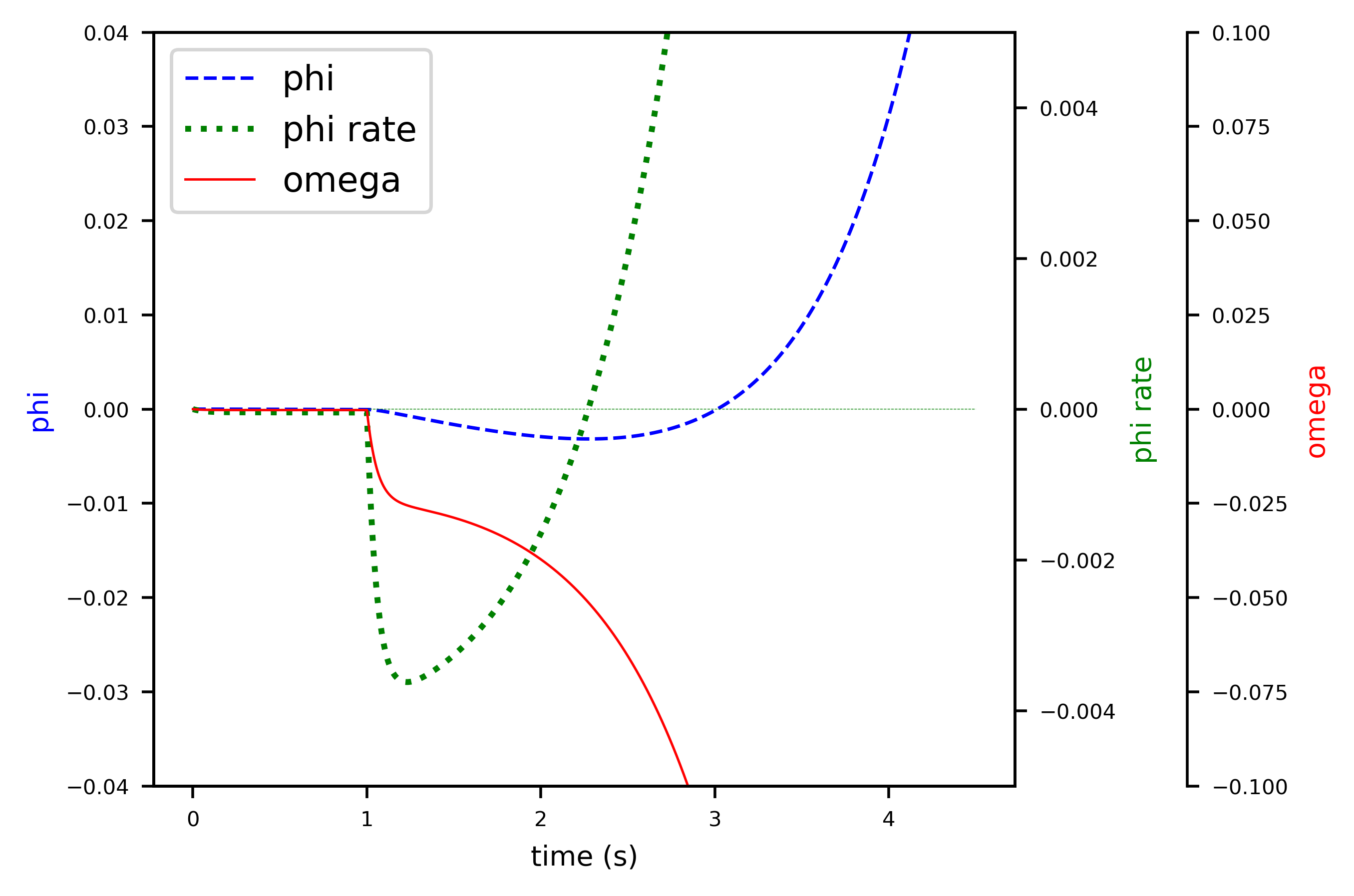}
\caption{an hypothetical situation where both conditions eq. \ref{eq:cond_nmpz} and eq. \ref{eq:cond_nmpz2} are true. 
At first, both $\dot \phi$ and $\omega$ decrease, but soon they both diverge because of the negative damping ($k_P$/$k_I$ corrections can delay this divergence but can not avoid it).}
\label{fig:both_nmpz}
\end{figure}

In order to complete the analysis of NMPZ phenomena related to FOWT system, 
a hypothetical situation where both conditions eq. \ref{eq:cond_nmpz} and eq. \ref{eq:cond_nmpz2} are true has been simulated and reported in Figure \ref{fig:both_nmpz}. 
At first, the dynamics are always the same: both $\dot \phi$ and $\omega$ decrease, 
but soon they both diverge because of the negative damping.
It is to be noted that, $k_P$/$k_I$ corrections (without compensations $\tau_g$ and $ k_{\beta} $) can delay this divergence but can not avoid it.

\begin{table}[ht]
\centering
\caption{Set of parameters to show the instability given by NMPZs of eq. \ref{eq:cond_nmpz} and eq. \ref{eq:cond_nmpz2}.}
\begin{tabular}[t]{ccc}
 \hline  
 $  $ & eq. \ref{eq:cond_nmpz} and eq. \ref{eq:cond_nmpz2} true \\ 
 \hline
 $\dTaV$ & $3105.0 ~ kN.s$ \\ 
 \hline
 $\dFaV$ & $293.0 ~ kN.s.m^{-1}$   \\ 
 \hline
 $\dTaOmg$ & $-51356.5 ~ kN.m.s.rad^{-1}$ \\ 
 \hline
 $\dFaOmg$ & $-7150.0 ~ kN.s.rad^{-1}$  \\
 \hline
 $\dTaBeta$ & $-148063.0 ~ kN.m.rad$ \\
 \hline
 $\dFaBeta$ & $-16543.6 ~ kN.rad$  \\
\hline
\end{tabular}
\label{tab:nmpz_concl}
\end{table}%

\subsection{Damping analysis}
\label{sec:damping}

In section \ref{NMPZ} the issue of NMPZ, ie. the issue of negative damping in the \textit{control/input side} of the equation, has been analysed.

On top of setting boundaries on gains $k_P$, $k_I$, $k_\beta$ and $k_{\tau_g}$ 
(cf. Section \ref{NMPZ}), 
this section analyses how they affect the explicit value of the damping. 
The goal is to optimize (or tune) the stability of $\omega$ and $\phi$ responses to an external ($v$ and $w$) disturbance. 
In other words, the goal is to obtain an explicit expression of the damping of the FOWT system with respect to the control parameters, $k_P$, $k_I$, $k_\beta$ and $k_{\tau_g}$, such that, for an imposed level of damping, one can obtain a value of the control parameters. 
This is a powerful result for the floating wind community and a novelty of this work with respect to the existing literature.

Considering the whole system, with both degrees of freedom $\omega$ and $\phi$ and their coupling, $\chi_A (s)$ of equation \ref{eq4} in the complex domain leads to the equation \ref{eq5}. 
The study of the damping is related to $\chi_A (s) = det(sI-A)$. 
The explicit form of $\chi_A$ is:

\begin{equation}
\label{eq10}
\chi_A (s)  = \chi_{rot}(s) \chi_{plt}(s) + \frac{N_g h_t}{J_r J_t} s 
\left[(k_p s + k_I) \dFaBeta h_t \dTaV + 
\left( \frac{J_r}{N_g} \dFaBeta (k_Ps + k_I) + \dFaOmg \right) \left(k_\beta \dTaBeta + k_{\tau_g} N_g \right) \right]
\end{equation}

where:

\begin{equation}
    \label{eq10b}
    \begin{split}
    \chi_{rot}(s) & = s^2 - \frac{N_g}{J_r} \dTaOmg s - \frac{N_g}{J_r} \dTaBeta (k_P s + k_I) \\
    \chi_{plt}(s) & = s^2 + \frac{1}{J_t} (D_t + h_t^2 \dFaV - k_\beta h_t \dFaBeta) s + \frac{K_t}{J_t}\\
    \end{split}
\end{equation}

The term in square parenthesis represents the coupling term between the dynamics of the platform ($\phi$) and the dynamics of the rotor ($\omega$).

In this coupled form, it is complicated to explicit the damping of the system. 
In the next paragraph, under some hypothesis, the coupled system can be separated in two second order systems, one related to the rotor dynamics $\omega$ and the other one related to the floating dynamics $\phi$. 
In particular, for the latter, it is possible to define a damping for the floating platform and obtain an explicit form for the compensation term $k_{\beta}$ related to the imposed damping.

\paragraph{Simplified analysis of rotor dynamics:}

Defining a damping coefficient (or a damping ratio) requires to reduce the global system to a second order oscillatory system. 
Equations \ref{eq4} couples rotor and platform pitch dynamics, they hence involve a $4^{th}$ order polynomial expression. 
In order to deal with rotor dynamics independently of the platform, it is supposed:

\begin{equation}
    \label{eq8}
    h_t \dot \phi \  \ll  \ v
\end{equation}
For large FOWT systems, this hypothesis is, generally, respected.   
It implies: 
\begin{equation}
\begin{split}
N_g k_{\tau_g} \dot \phi & \ll \dTaV v \\
\dTaBeta k_\beta \dot \phi & \ll \dTaV v \\ 
\end{split}
\end{equation}

Under such assumption, the linear form of equation \ref{eq1} becomes:

\begin{equation}
    \dot \omega = \frac{N_g}{J_r} \left(\dTaOmg \omega + 
            \dTaV v + 
            \dTaBeta \beta - \tau_g \right) \tag{1''}
\end{equation}

and the control is described by the PI controller: $\dot \beta = k_P \dot \omega + k_I \omega$, 
such that the resulting Laplace transform equation is 
\begin{equation}
\underline \omega (s) = G_{rot} (s) \underline  v (s)
\end{equation}
where, considering a $k_I > 0$,
\begin{equation}
    \label{eq9}
    \begin{split}
    G_{rot}(s&) =\frac{\dTaV s}{s^2 - \frac{N_g}{J_r} \dTaOmg s - \frac{N_g}{J_r} \dTaBeta (k_P s + k_I)}, \\
    i.e. \quad G_{rot}(j\nu&) =\frac{1}{1 + \frac{j}{2 \zeta_{rot}} \left(\frac{\nu}{\nu_{rot}} - \frac{\nu_{rot}}{\nu} \right)} \frac{- \dTaV}{\frac{N_g}{J_r} \dTaOmg + \frac{N_g}{J_r} \dTaBeta k_P} \\
    \end{split}
\end{equation}
with:
\begin{equation}
\label{eq:PI_tuning}
     \nu_{rot} = \sqrt{- \frac{N_g}{J_r} \dTaBeta k_I} \; , \quad \zeta_{rot} = - \frac{\frac{N_g}{J_r} \dTaOmg + \frac{N_g}{J_r} \dTaBeta k_P}{2 \sqrt{-\frac{N_g}{J_r} \dTaBeta k_I}}
\end{equation}

Thus, when all interactions with platform pitch are neglected, the rotor behaves like a second order oscillatory system. The corresponding filter $G_{rot}$ is a second order band-pass filter with cutoff angular frequency $\nu_{rot}$. In case $k_I \leq 0$, $\nu_{rot}$ and $\zeta_{rot}$ would be imaginary according to the formulas above. $G_{rot}$ is no longer a band-pass filter.

The above formulas enable one to obtain explicitly $k_I$ and $k_P$. 

They are well known: several controllers, such as ROSCO \cite{rosco}, suggest to define: 
\begin{equation}
\left| k_I \right| = \left| \frac{\nu_{rot}^2}{\frac{N_g}{J_r} \dTaBeta} \right|  \quad  \text{and}  \quad \left| k_P \right| = \left| \frac{\left( \frac{N_g}{J_r} \dTaOmg + 2 \zeta_{rot} \nu_{rot} \right)}{\frac{N_g}{J_r} \dTaBeta} \right| 
\end{equation}

\paragraph{Simplified analysis of platform dynamics:}

Similarly as what is done in the previous paragraph, here the global system \ref{eq1} is reduced to a second order oscillatory system that allows us to have a better understanding of platform dynamics.

Considering $k_P = k_I = 0$ and assuming: 
\begin{equation}
\dFaOmg \omega << \dFaV v_r + \dFaBeta \beta
\end{equation}
The latter is the condition to decouple the global system. 
It enables to consider $\phi$ response independently of $\omega$, 
and as a second order oscillatory system's degree of freedom. 
The resulting Laplace transform equation is:
\begin{equation}
\label{eq:plt_dyn}
    \begin{pmatrix}
    \underline \phi \\
    \underline {\dot \phi} \\
    \end{pmatrix}
(s) = G_{plt}(s) \underline  u_d (s)
\end{equation}
$u_d = \begin{pmatrix}
    v \\
    w \\
    \end{pmatrix}$: 
the input array is reduced because only the damping in the \textit{output side} is analysed and it is not necessary for this to consider any additional control input.
\begin{equation}
G_{plt}(s) = (sI - A_{plt})^{-1} B_d 
\end{equation}
\begin{equation}
A_{plt} = 
\begin{bmatrix}
0                 & 1                                                           \\
- \frac{K_t}{J_t} & \frac{1}{J_t} (-D_t - h_t^2 \dFaV + k_\beta h_t \dFaBeta)   \\
\end{bmatrix}
\end{equation}
$A_{plt}$ is the right part of $A$ defined in \ref{SS description} as $B_d$.

Looking at the $\phi$ degree of freedom, eq. \ref{eq:plt_dyn} gives:
\begin{equation}
\underline  \phi (s) = G_{plt, 1, 1} (s) \underline  v (s) + G_{plt, 1, 2} (s) \underline  w (s) ,
\end{equation}
with: 
\begin{equation}
\left( G_{plt, 1, 1}, G_{plt, 1, 2} \right) (s)  = 
\begin{pmatrix}
    \frac{\frac{h_t}{J_t} \dFaV}{s^2 + \frac{1}{J_t} (D_t + h_t^2 \dFaV - k_\beta h_t \dFaBeta) s + \frac{K_t}{J_t}} ,  \frac{\frac{1}{J_t} \dTaW}{s^2 + \frac{1}{J_t} (D_t + h_t^2 \dFaV - k_\beta h_t \dFaBeta) s + \frac{K_t}{J_t}}
\end{pmatrix},
\end{equation}
i.e.:
\begin{equation}
\left( G_{plt, 1, 1}, G_{plt, 1, 2} \right) (j \nu)  = \frac{1}{1 - \left( \frac{\nu}{\nu_{plt}} \right)^2 + 2 j \zeta_{plt} \frac{\nu}{\nu_{plt}}}
\begin{pmatrix}
 \frac{h_t}{K_t} \dFaV ,  \frac{1}{K_t} \dTaW\\
\end{pmatrix}  , 
\end{equation}

\begin{equation}
\label{eq:plt_dyn_param}
\nu_{plt} = \sqrt{\frac{K_t}{J_t}} \; , \qquad \zeta_{plt} = \frac{1}{2 \sqrt{K_t J_t}} \left(D_t + h_t^2 \dFaV - k_\beta h_t \dFaBeta \right)
\end{equation}

Thus, when all interactions with rotor dynamics are neglected, the platform behaves like a second order oscillatory system. 
The corresponding filter $G_{plt}$ is a second order low-pass filter with cutoff angular frequency defined by $\nu_{plt}$ and damping ratio defined by $\zeta_{plt}$.

\subsection{Artificial damping of the platform : $\zeta_{plt}$-fixed strategy}

\label{Ptfm}

By knowing the features of the FOWT, one can impose a given level of damping and obtain an explicit expression for the $k_{\beta}$:

\begin{equation}
\label{k_b_strategy}
k_\beta = \frac{1}{h_t \dFaBeta} \left(D_t + h_t^2 \dFaV - 2 \sqrt{K_t J_t} \zeta_{plt} \right) 
\end{equation}

The strategy is such that $k_\beta$ is a negative number instead of what is proposed in \cite{rosco, Stockhouse}. 
In those articles, $\beta_{comp} = k_\beta \dot \phi$, a positive value, is introduced in order to erase at first order the coupling between platform and rotor dynamics, and therefore is defined by : 
\begin{equation}
    k_\beta = h_t \dTaV / \dTaBeta
\end{equation}

\paragraph{Expected effect on platform dynamics}

\begin{figure}[H]
\centering
\includegraphics[scale=0.4]{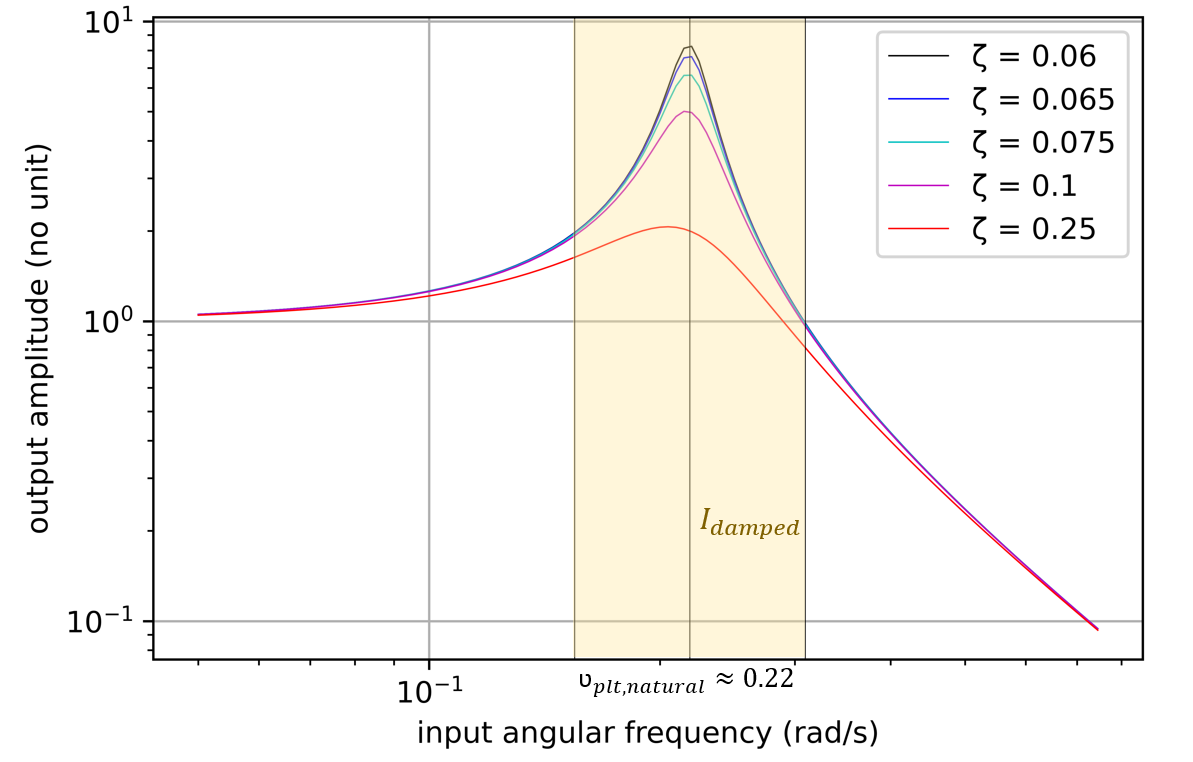}
\caption{Bode diagram of second order low-pass filter $G_{plt}$}
\label{fig:Bode}
\end{figure}

Figure \ref{fig:Bode} shows the second order low-pass filter $G_{plt}$. In other words, it is how $\zeta_{plt}$ value can affect the damping of platform oscillations.

It can be observed that $\zeta_{plt}$ has a significant effect on the damping of platform oscillations only for angular frequencies $\nu \approx \nu_{plt, natural}$. 
Yellow vertical band in figure \ref{fig:Bode} shows the interval of angular frequencies $I_{damped}$, arbitrarily defined by:
\begin{equation}
I_{damped} = \left[\frac{\nu_{plt, natural}}{\sqrt{2}} ~, ~ \sqrt{2} \nu_{plt, natural} \right]
\end{equation}
that are directly damped when $\zeta_{plt}$ increases.
Therefore, it is to be expected that $\zeta_{plt}$-fixed strategy will be well fit to reduce platform motion and tower loads when their variations happen at an angular frequency $\nu \in I_{damped}$. 

\paragraph{Expected effect on rotor dynamics}

In this part the first order effect of $\zeta_{plt}$-fixed strategy on the rotor dynamics is analysed.
The state space representation of the FOWT dynamics is given by equation \ref{eq4}.
By considering $0$ input (ie. $u_{c, ol} = u_d = 0$), this leads to the following linear equation, truncated at first order:
\begin{equation}
    \label{fig:first_order}
    \dot \omega = \ddot \theta = A_{2, 1} \theta + A_{2, 2} \dot \theta + A_{2, 4} \dot \phi,
\end{equation}

where

\begin{equation}
    \label{fig:first_order}
    \begin{split}
    A_{2, 4} & =\frac{N_g}{J_r} \left(-h_t \dTaV + k_\beta \dTaBeta - k_{\tau_g} N_g \right)\\
    & = \frac{N_g}{J_r} \left(-h_t \dTaV + \frac{\dTaBeta}{h_t \dFaBeta} \left(D_t + h_t^2 \dFaV - 2 \sqrt{K_t J_t} \zeta_{plt} \right) - k_{\tau_g} N_g \right). \\
    \end{split}
\end{equation}

Moreover, the following inequalities are verified for an above-rated operating point: 
\begin{equation}
    \label{fig:inequalities}
    \begin{split}
    h_t \dTaV + k_{\tau_g} N_g & \geq 0 \\
    \frac{\dTaBeta}{\dFaBeta} & > 0 \\
    \frac{1}{2 \sqrt{K_t J_t}} \left(D_t + h_t^2 \dFaV \right) & \leq \zeta_{plt} \\
    \end{split}
\end{equation}

First inequality comes from \ref{ktau_g} (notice that in section \ref{numerical}, it is considered $\tau_g = 0$). Third inequality is a consequence of the assumption that $\zeta_{plt}$-fixed strategy aims at increasing the damping of the platform.
This implies that:
\begin{equation}
\frac{\partial}{\partial_{\zeta_{plt}}} \left| A_{2, 4} \right| > 0
\end{equation}

meaning that the first order coupling between platform dynamics and rotor dynamics will increase when $\zeta_{plt}$ increases if $\zeta_{plt}$-fixed strategy is applied. 
Thus, if the characteristic time of platform dynamics is small enough, the equation truncated at first order is valid and it is to be expected that, at least for some tunings of the PI controler, $\zeta_{plt}$-fixed strategy would increase rotor speed variations.

\section{Numerical tests with time domain simulations}

In this section, it is analysed how the new control strategy described in the previous section affects platform and rotor dynamics, and especially the impact on tower loads and rotational speed. 
The \textit{reference} is the control strategy one without compensation, with $k_\beta = 0$. 

The $\zeta_{plt}$-fixed strategy has been implemented in the ROSCO environment \cite{ROSCO_toolbox_2021}, replacing the existing pitch control. 
The rest of the controller remains basically the same. 
In the next, $\zeta_{plt}$-fixed strategy is compared to the one without compensation, named \textit{reference}. 
The only difference between the two terms of comparison concerns the $k_\beta$. For the $\zeta_{plt}$-fixed strategy, it is given by \ref{eq:plt_dyn_param}. For the reference, it is $k_\beta = 0$.

\label{numerical}

\subsection{Test cases}

The simulation tool used is OpenFast v2.4.0 (https://github.com/OpenFAST/openfast) and the FOWT model considered is the IEA 15 MW wind turbine mounted over the UMaine VolturnUS-S semi-submersible floater \cite{UMaine}. 
Initially simple constant wind and monochromatic waves are testes in order to verify the analytical developments of the previous section. 
Then a test case more representative of the industrial design of FOWT is considered by testing a DLC1.1. 
For the latter, simulation consider only $1$ seed of $3600$ seconds with aligned wind and irregular waves. For this time simulation, this is statistically equivalent to $600$ seconds and $6$ seeds.

\subsection{Tuning of the PI controller}

\label{sec:tuningPI}

Values of $k_P$, $k_I$, $k_\beta$ are continuously updated following the explicit expression given in equations \ref{eq:PI_tuning} and \ref{eq:plt_dyn_param}, then low-pass filtered. 
This means that globally, for each of these test cases and each set of parameters, $k_P$, $k_I$, $k_\beta$ have almost fixed values. 
After some tests, for all the considered simulations, $\zeta_{rot} = 0.6$ and $\nu_{rot} = 0.01$ are chosen for the PI controller's tuning. 
This choice ensures that most of the wave spectrum (which peaks at $T \approx 11 ~ s$, ie. $\nu \approx 0.57 ~ rad s^{-1}$) and platform dynamics natural angular frequency ($\nu_{plt} \approx 0.22 ~ rad.s^{-1}$) fall outside of $G_{rot}$ pass-band.
In the next, several values of platform damping are analyzed for the $\zeta_{plt}$-fixed strategy and compared to the reference strategy for different test case.

\subsection{Still wind and monochromatic wave}
For the still wind and monochromatic wave condition, 
two $\zeta_{plt}$ are tested: $\zeta_{plt} = 0.1$ and $\zeta_{plt} = 0.25$. The corresponding quality factors are: $Q = 5$ and $Q = 2$, respectively. 
Thus, the platform is expected to behave like an under-damped second order oscillatory system. 
Table \ref{tab:test_cases} states external conditions for test cases with still wind and monochromatic wave. 
Hereafter, results are plotted over time are drawn for a $100$ seconds time interval in a simulation on a long period of time, so that the operating point is reached. When necessary for a better understanding, results are reported for a longer interval. 

\begin{table}[ht]
\centering
\caption{Environmental conditions for the numerical test cases.}
\begin{tabular}[t]{ p{3.0cm}  p{3.0cm} p{3.0cm} p{3.0cm}  }
 \hline
 case & wind speed $V$ ($ms^{-1}$) & wave period $Tp$ ($s$) & wave height $H_w$ ($m$)\\
 \hline
 (1) & 11 & 11 & 1.5 \\
 (2) & 11 & 28.75 & 1.5 \\
 (3) & 22 & 11 & 1.5 \\
 (4) & 22 & 28.75 & 1.5 \\
\hline
\end{tabular}
\label{tab:test_cases}
\end{table}

Table \ref{tab:k_beta} gives the mean value of $k_\beta$ for test cases (1) to (4). Cases (1), (2) and cases (3), (4) are gathered together as they use the same mean value of $k_\beta$. \\
\begin{table}[ht]
\label{tab:k_beta}
\caption{compensation gain ($k_\beta$).}
\centering
\begin{tabular}[t]{ p{2.5cm}  p{2.5cm}  p{2.5cm}  p{2.5cm}  }
 \hline
 case &  $\zeta_{plt} = 0.10$ &  $\zeta_{plt} = 0.25$ & reference \\
 \hline
(1) and (2)
  & $k_\beta = -8.6$ &  $k_\beta = -42.7$ & $k_\beta = 0.0$\\
 (3) and (4)
 &  $k_\beta = -7.4$  &  $k_\beta = -34.8$ &  $k_\beta = 0.0$\\
\hline
\end{tabular}
\label{tab:k_beta}
 \end{table}

\begin{figure}[H]
\centering
\includegraphics[scale=0.5]{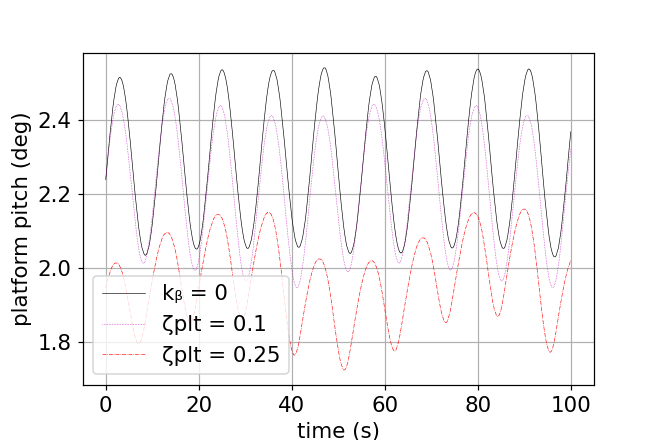}
\includegraphics[scale=0.5]{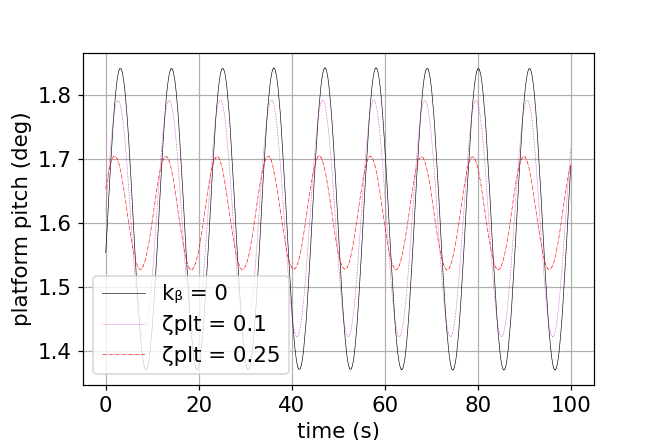}
\caption{Platform pitch $\Phi$ (deg) for a monochromatic wave of period 11s (test case (1) on the left and (3) on the right}
\label{fig:PtfmPitch1}
\end{figure}

\begin{figure}[H]
\centering
\includegraphics[scale=0.5]{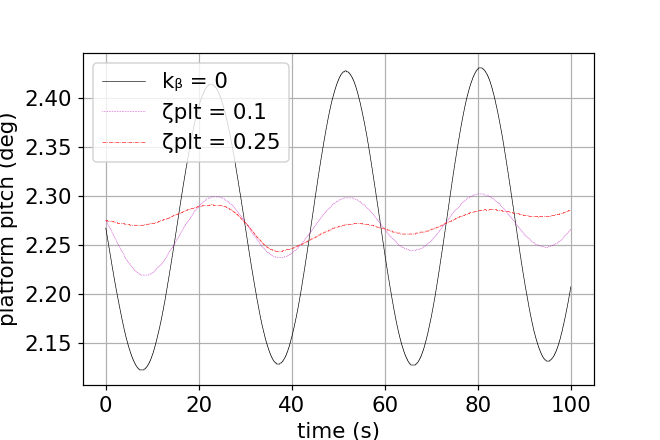}
\includegraphics[scale=0.5]{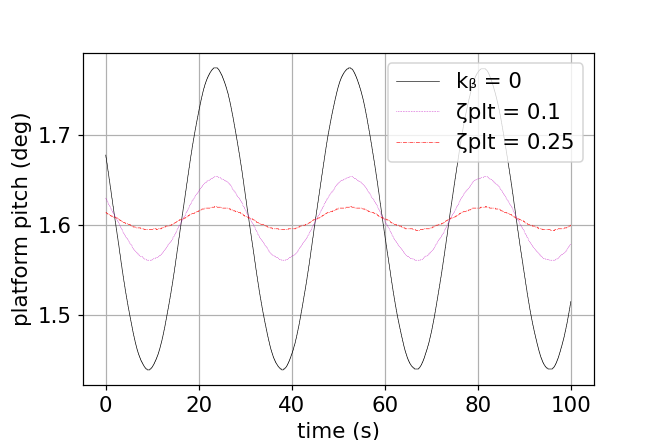}
\caption{Platform pitch $\Phi$ (deg) for a monochromatic wave of period 28.75s (test case (2) on the left and (4) on the right}
\label{fig:PtfmPitch2}
\end{figure}

\begin{figure}[H]
\centering
\includegraphics[scale=0.5]{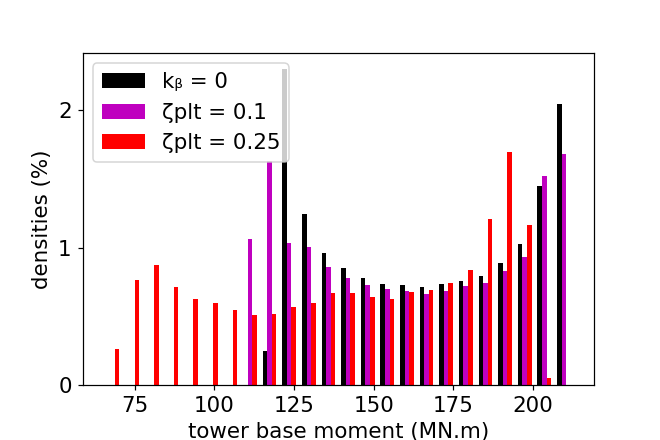}
\includegraphics[scale=0.5]{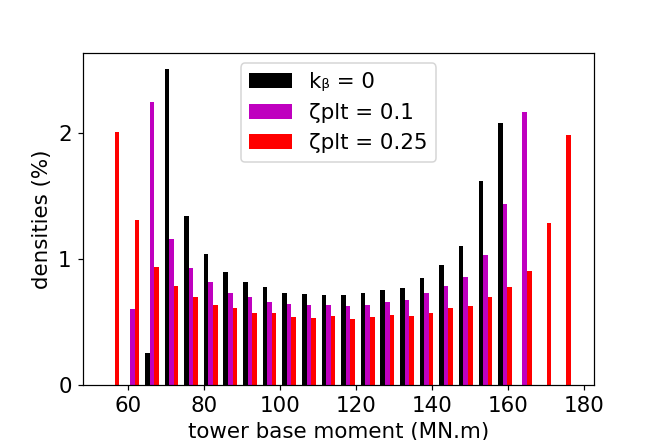}
\caption{Tower base moment (MN.m) densities for a monochromatic wave of period 11s (test case (1) on the left and (3) on the right}
\label{fig:moment1}
\end{figure}

\begin{figure}[H]
\centering
\includegraphics[scale=0.5]{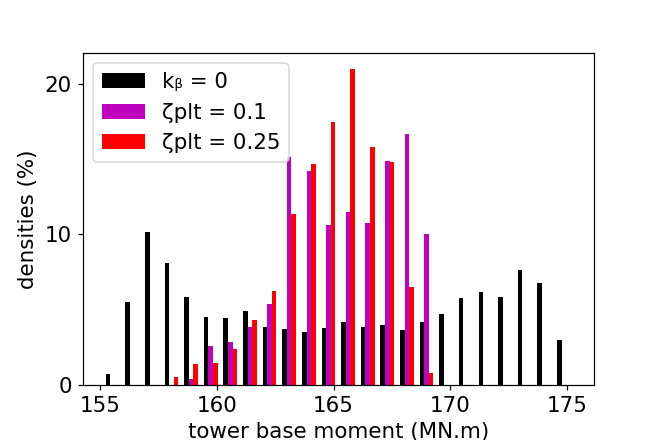}
\includegraphics[scale=0.5]{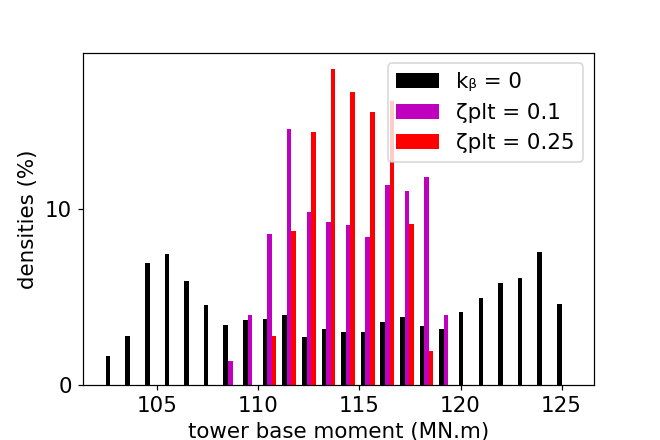}
\caption{Tower base moment (MN.m) densities for a monochromatic wave of period 28.75s (test case (2) on the left and (4) on the right}
\label{fig:moment2}
\end{figure}

Figures \ref{fig:PtfmPitch1} and \ref{fig:PtfmPitch2} show the forced oscillations of the platform when it is subjected to a monochromatic wave of period $11s$ (which corresponds to the realistic fundamental period of a wave) and $28.75s$ (which is the natural period of the platform). 
As shown by the analytical development, increasing $\zeta_{plt}$ reduces platform oscillations, especially when the wave period is close to the natural period of the platform (see diagram \ref{fig:Bode}). 
This is shown in Figures \ref{fig:moment1} and \ref{fig:moment2} where the density of occurrence of each value of tower bottom reaction moment is reported. 
For the monochromatic wave with period $11s$ the damping of the $\zeta_{plt}$-fixed strategy is less evident. 
More floating wind test cases are reported in the next section.  

\begin{figure}[H]
\centering
\includegraphics[scale=0.5]{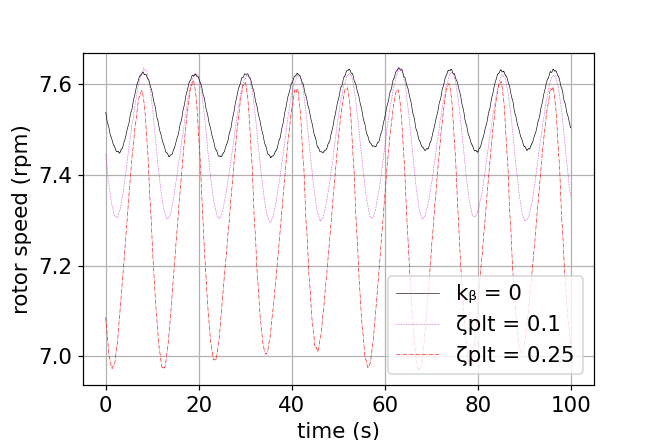}
\includegraphics[scale=0.5]{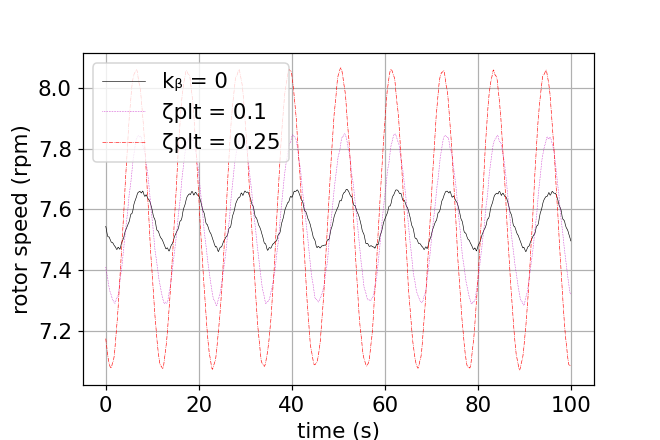}
\caption{Rotor speed (rpm) evolution over time for test cases (1) on the left and (2) on the right}
\label{fig:rotSpeed1}
\end{figure}

\begin{figure}[H]
\centering
\includegraphics[scale=0.5]{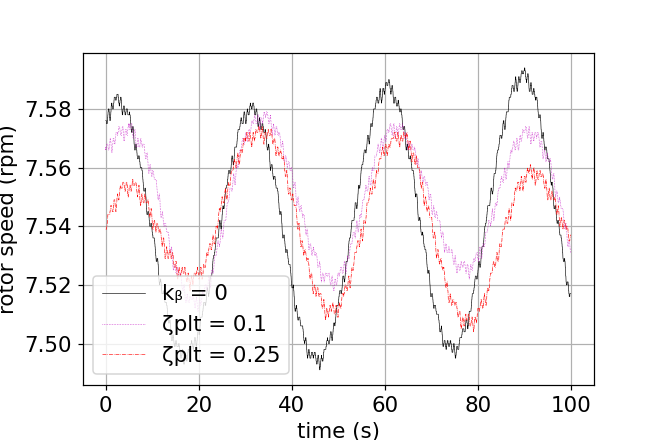}
\includegraphics[scale=0.5]{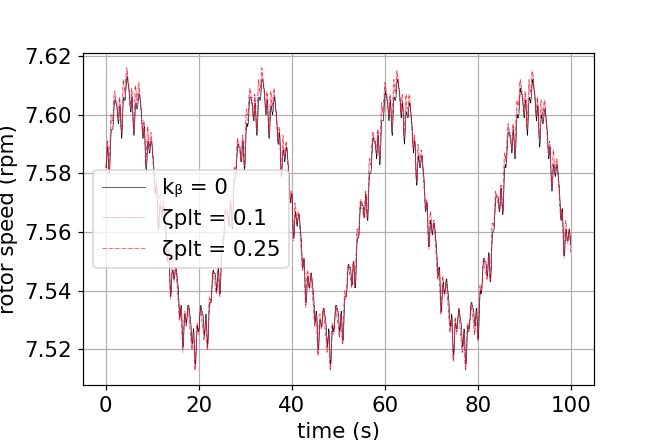}
\caption{Rotor speed (rpm) evolution over time for test cases (3) on the left and (4) on the right}
\label{fig:rotSpeed2}
\end{figure}

$\zeta_{plt}$-fixed strategy's effect on rotor dynamics is not easily described by a second order linear equation: it involves the coupling between platform and rotor dynamics, which was analysed at first order in \ref{Ptfm}. 
From this analysis, $\zeta_{plt}$-fixed strategy was expected to increase the coupling between platform and rotor dynamics for a short characteristic time. 
In Figure \ref{fig:rotSpeed1} and \ref{fig:rotSpeed2} it can be observed that for a short characteristic time ($11s$) $\zeta_{plt}$-fixed strategy increases rotor speed variations, but not for a longer characteristic time, such as $28.75s$ (for which it behaves slightly better than $k_\beta=0$ strategy). 

To conclude this part of the tests: 
$\beta$-compensation strategies perform very differently depending on the oscillatory frequency of the platform:
\begin{itemize}
\item For angular frequencies $\nu \in I_{damped} = \left[\frac{\nu_{plt, natural}}{\sqrt{2}} ~, ~ \sqrt{2} \nu_{plt, natural} \right]$), 
$\zeta_{plt}$-fixed strategy is very effective when it comes to the damping of platform oscillations, 
as seen in paragraph \ref{Ptfm} (cf. figure \ref{fig:Bode}). 
The tests highlight that $\zeta_{plt}$-fixed strategy is reducing both tower loads and rotor speed variations in turbulent wind conditions.
\item For angular frequencies outside the previous set, $\zeta_{plt}$-fixed strategy is less effective for damping platform oscillations. Tower loads reduction by $\zeta_{plt}$-fixed strategy is therefore barely visible, whereas rotor speed variations are actually amplified, especially when comparing this strategy to reference strategy.
\end{itemize}

\subsection{DLC1.2 tests} 

Tests presented hereafter are more representative of what is done during design or verification of offshore wind structure. As already done for the previous test cases, the $\zeta_{plt}$-fixed strategy is compared to the reference strategy, $k_{\beta} = 0$. 
They are inspired by the DLC 1.2, for normal power production in normal turbulence and normal sea state, as described in IEC standards. This kind of load case aim at assessing the fatigue design criteria. 
Kaimal's turbulence model is considered following IEC 61400 v.3 for a Wind Turbine of turbulence type B, for average wind speeds from $12 ms^{-1}$ to $24 ms^{-1}$, as described in Table \ref{tab:env_cond}. The wind box is generated by TurbSim tool developed by NREL.  
For the waves, JONSWAP distributions are considered with significant wave height $Hs=1.5 m$, wave period $Tp=11.0 s$ and $\gamma=2.0$. 
Wind and waves are considered aligned in the same direction. 

For those test cases, the level of damping imposed to the platform is $\zeta_{plt}$-fixed = $0.1$. 
This value is found to be the interesting to be tested for this floater and WTG configuration. 
Other tests with higher values of imposed damping show less interesting results. 
The choice of the right $\zeta_{plt}$ for each FOWT system is important and demands for some iterations before concluding.

\begin{table}[ht]
\centering
\caption{Environmental conditions for DLC 1.2.}
\label{tab:env_cond}
\begin{tabular}[t]{p{2.0cm} p{2.5cm} p{3.5cm}  p{1.0cm} p{1.5cm} p{1.0cm} p{1.5cm}  p{2.0cm}  }
 \hline
Time sim [$s$] & w.speed [$m.s^{-1}$] & w. condition & $Tp$ [$s$] &  $H_s$ [$m$] & $ \gamma $ & waves dir. \\
 \hline
$3600$ & $12.0 - 24.0$ & Normal turbulence B  & $11.0$ & $1.5$ & $2.0$ & co-linear \\
\hline
\end{tabular}
\label{tab:test_cases}
\end{table}

\begin{figure}[!ht]
\centering
\includegraphics[scale=0.6]{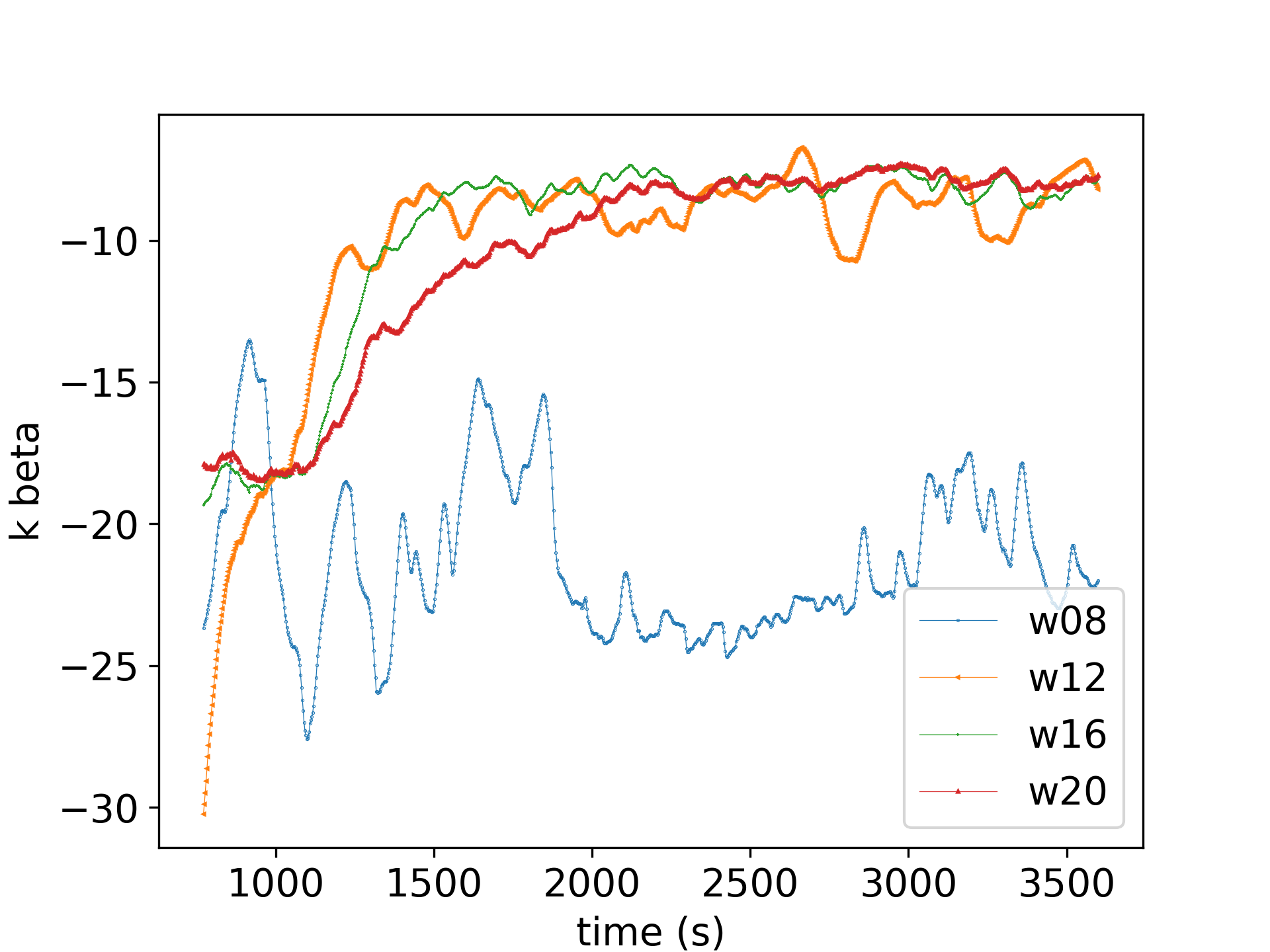}
\caption{evolution of $k_\beta$ for some of the simulations.}
\label{fig:kb_evol}
\end{figure}

Figure \ref{fig:kb_evol} shows how $k_\beta$ evolves for some cases of the simulation pool.
The below-rated behaviour is more dynamic than the over-rated, where the floating feedback is more stable. 
As remarked in \ref{Ptfm}, the $k \beta$ for the $\zeta_{plt}$-fixed is a negative value. 
This is the novelty of this strategy with respect to other floating feedback strategies, as the one proposed in \cite{rosco}. 

\begin{figure}[!ht]
\centering
\includegraphics[scale=0.45]{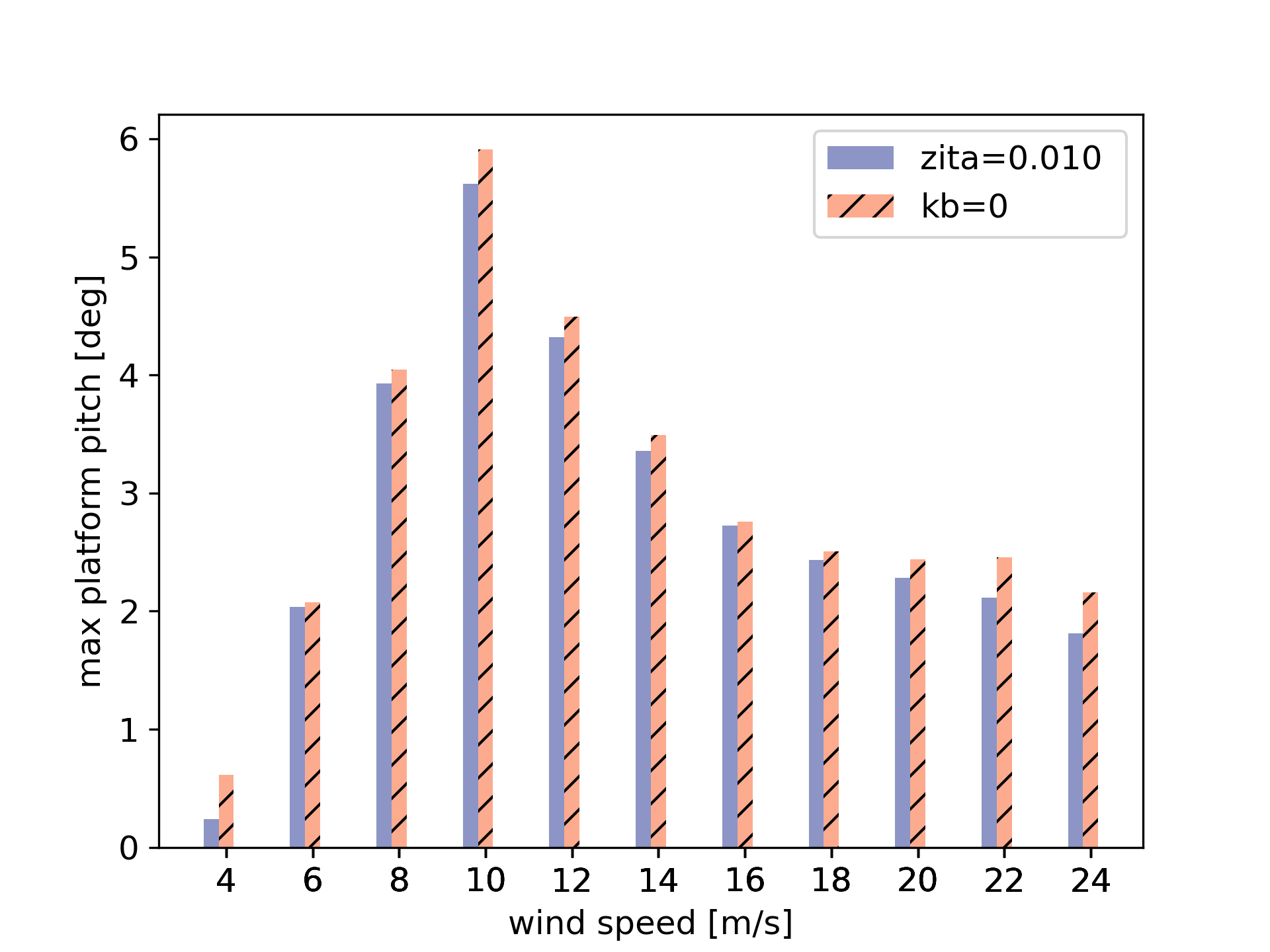}
\includegraphics[scale=0.45]{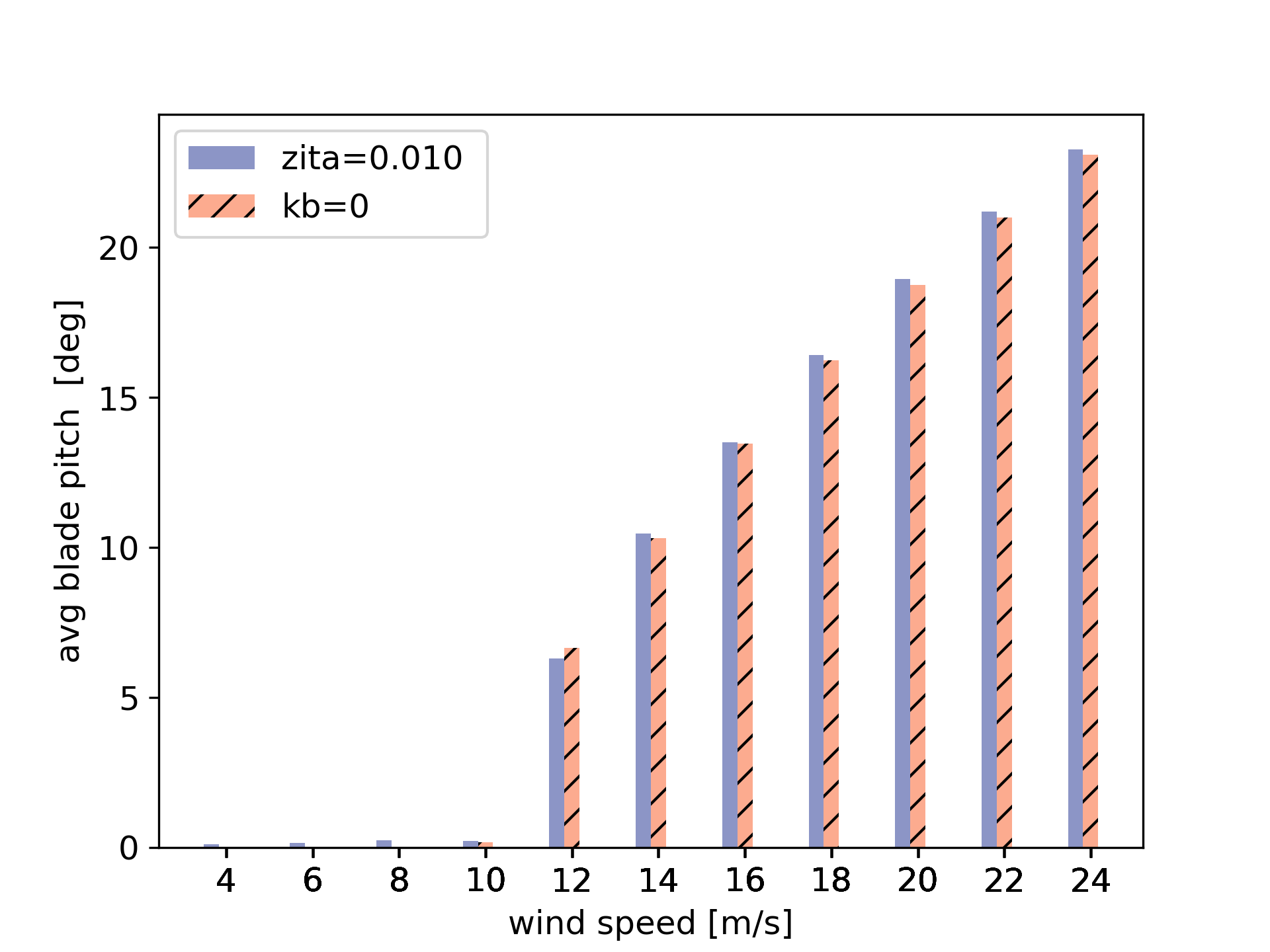}
\includegraphics[scale=0.45]{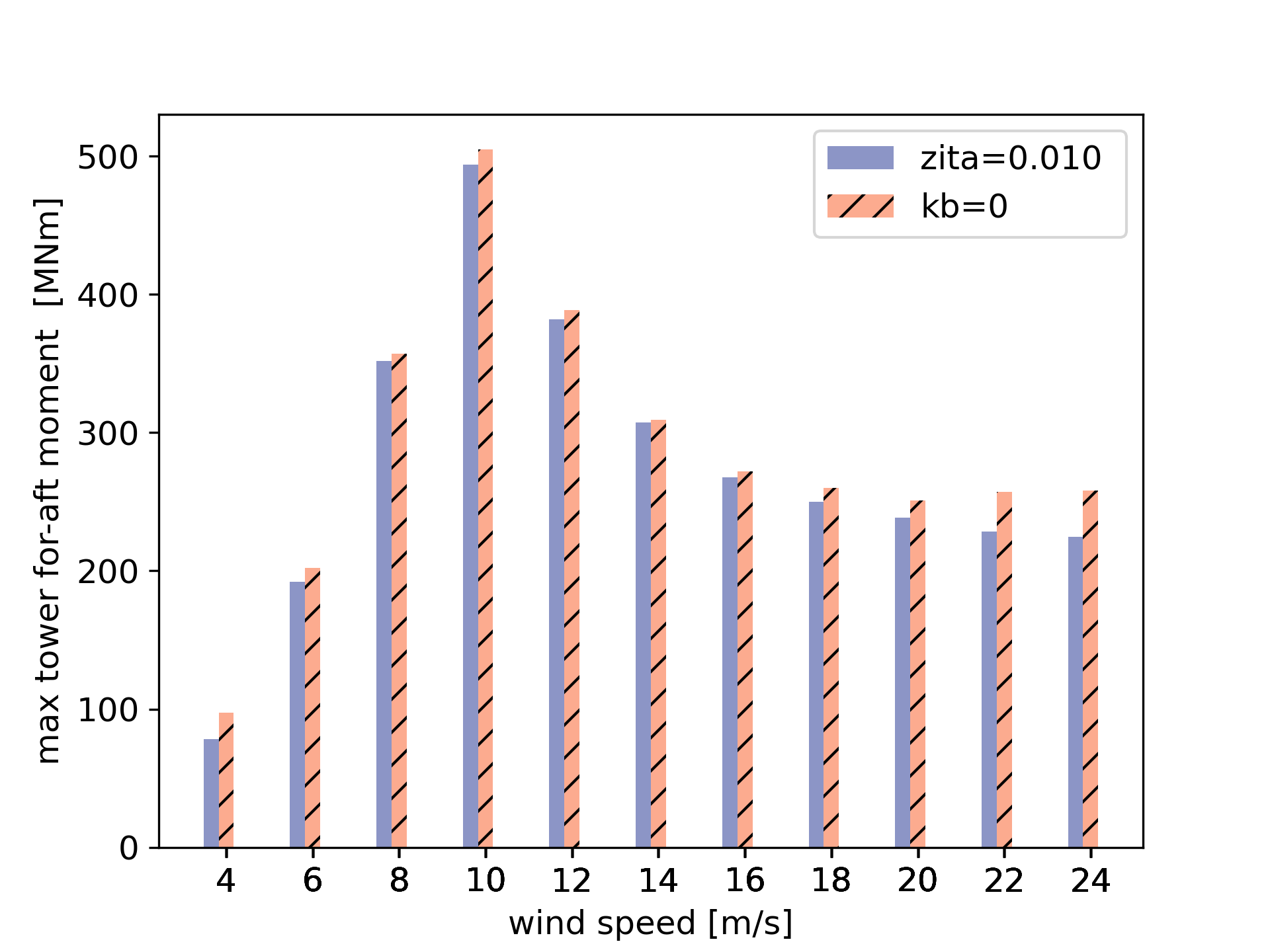}
\includegraphics[scale=0.45]{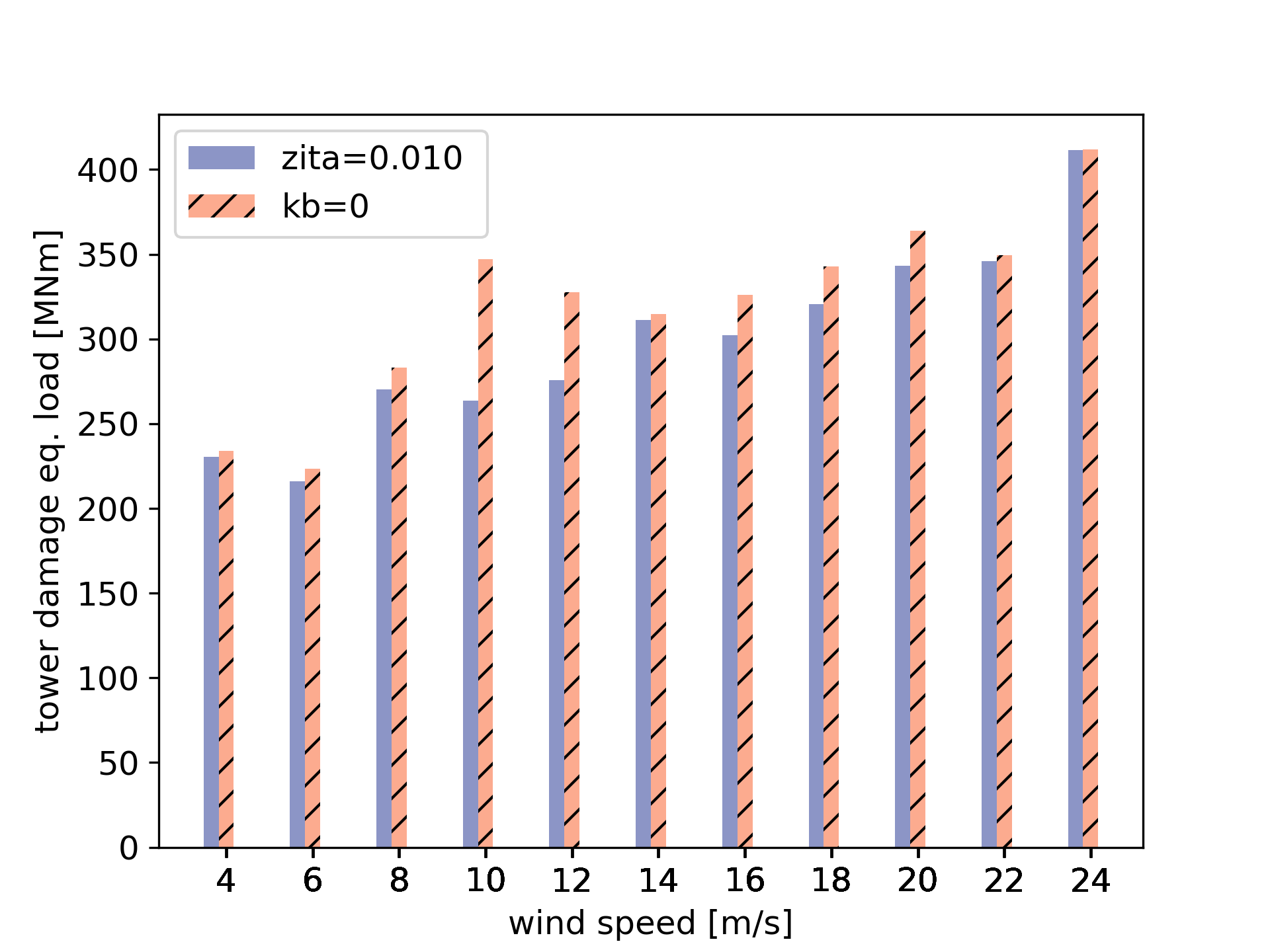}
\includegraphics[scale=0.45]{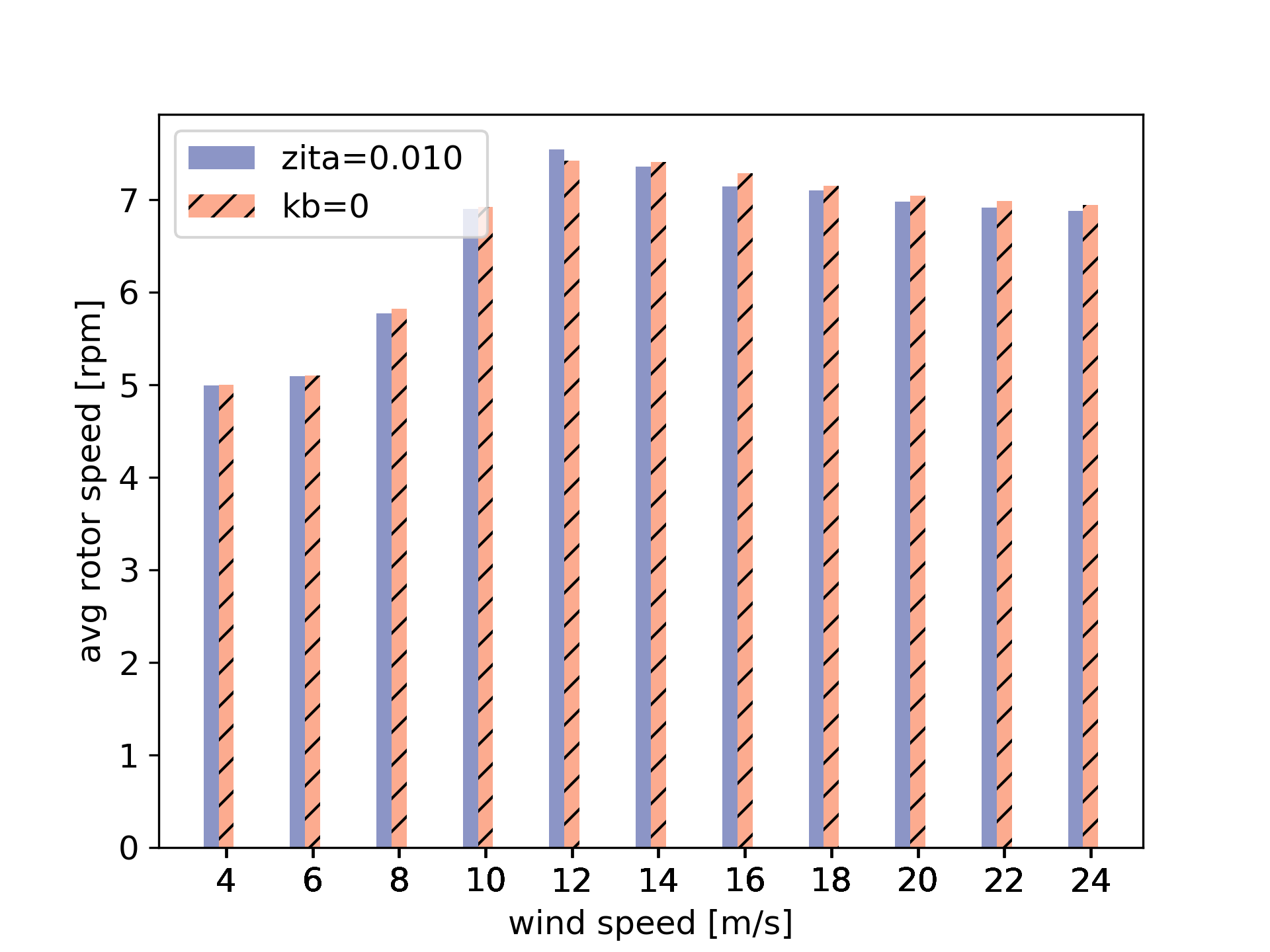}
\includegraphics[scale=0.45]{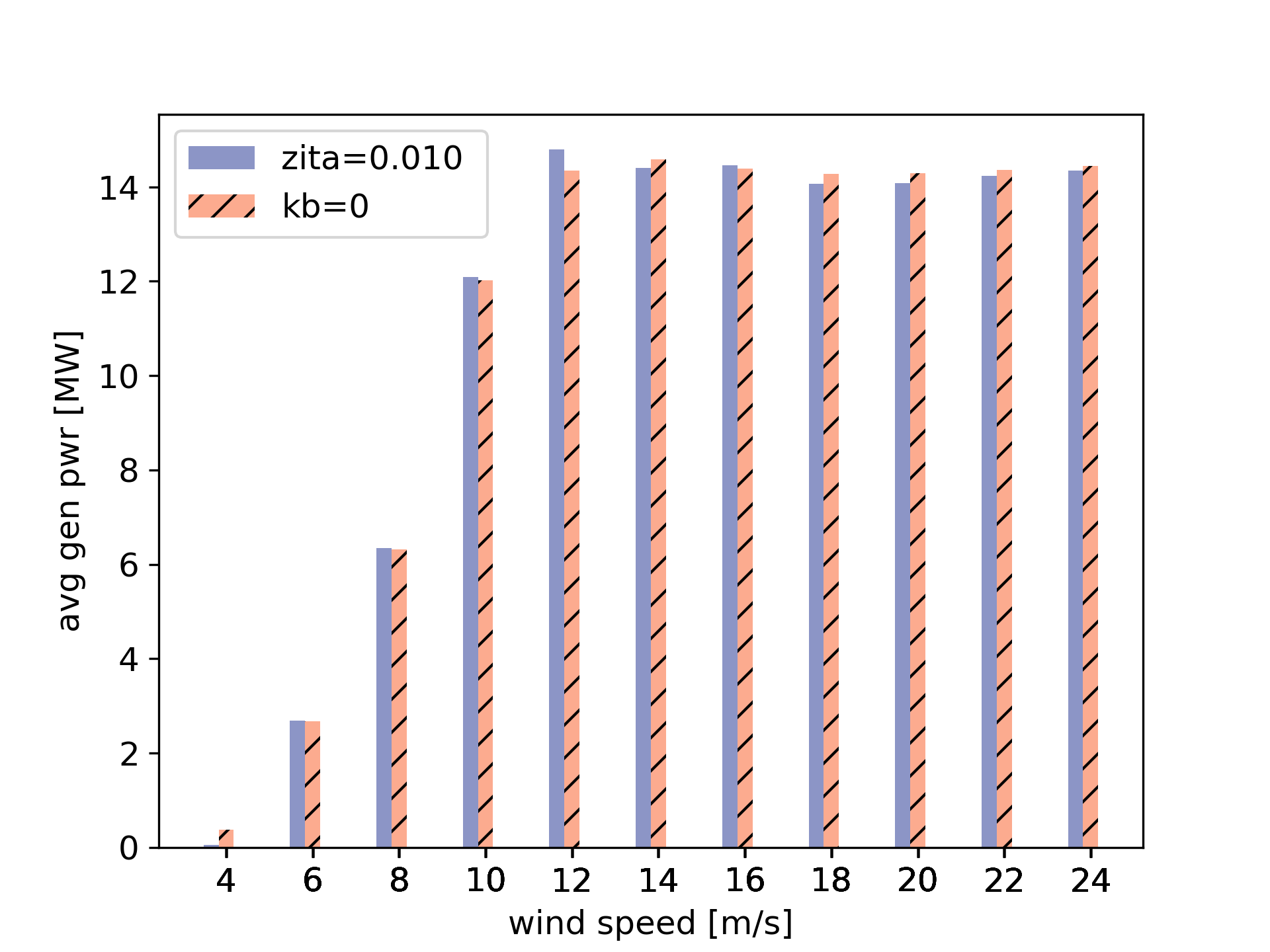}
\caption{Comparison results for the DLC1.2 for the UMaine floater with IEA15MW WTG. 
The imposed level of damping in the platform dynamics for the $\zeta_{plt}$-fixed strategy is $0.10$. 
Output show statistics for platform pitch; blade pitch; Tower bending moment, max and damage equivalent load; rotor speed and generator power. }
\label{fig:results_sim_turb}
\end{figure}

From the time series of the tower bottom moment, a rainflow algorithm is used to count the cycles following ASME normative for steel structures \cite{ASME}. 
The DEL is obtained by using a Wohler's curve with a single slope of exponent $m=3.0$. 
Platform pitch, power, rotor speed, blade pitch, tower load and tower DEL results of the simulations for the comparison are resumed in Figure \ref{fig:results_sim_turb}. 
The $\zeta_{plt}$-fixed strategy reduces platform pitch motion for all wind speeds. 
As expected this is done by the coupling with the rotor and the use of the blade pitch. 
In fact, blade pitch is more used by considering the $\zeta_{plt}$-fixed strategy. 
This was expected by the analytical development. 
As shown, the coupling between platform pitch and rotor dynamics is increased. 
The increase in pitch utilisation explains the slight decrease in rotor speed for the above rated wind speeds.
The generator power is also affected, slightly increased for lower wind speeds and slightly decreased for higher wind speed (around $+/- 1\%$). 

Loads in the tower are reduced and fatigue Design Equivalent Load (DEL) also. 
This is remarkable for around rated speed. For high wind speeds, the gain is less evident.  
In average there is a gain around $15 \%$ of the DEL. 
Table \ref{tab:res_10ms} shows, for the $10$ $m s^{-1}$ case, a deeper comparison by reporting the statistics of the quantities of interest extracted from this simulation. 
The difference in the minimum value of the moment at the tower base is remarkable: passing from $80.97 $ $MNm$ for the reference up to $165.40$ $MNm$ for the $\zeta_{plt}$-fixed strategy. 
This is also clear in the difference of the standard deviations. 
The amplitudes of the oscillations of this load are reduced.

\begin{figure}[!ht]
\centering
\includegraphics[scale=0.6]{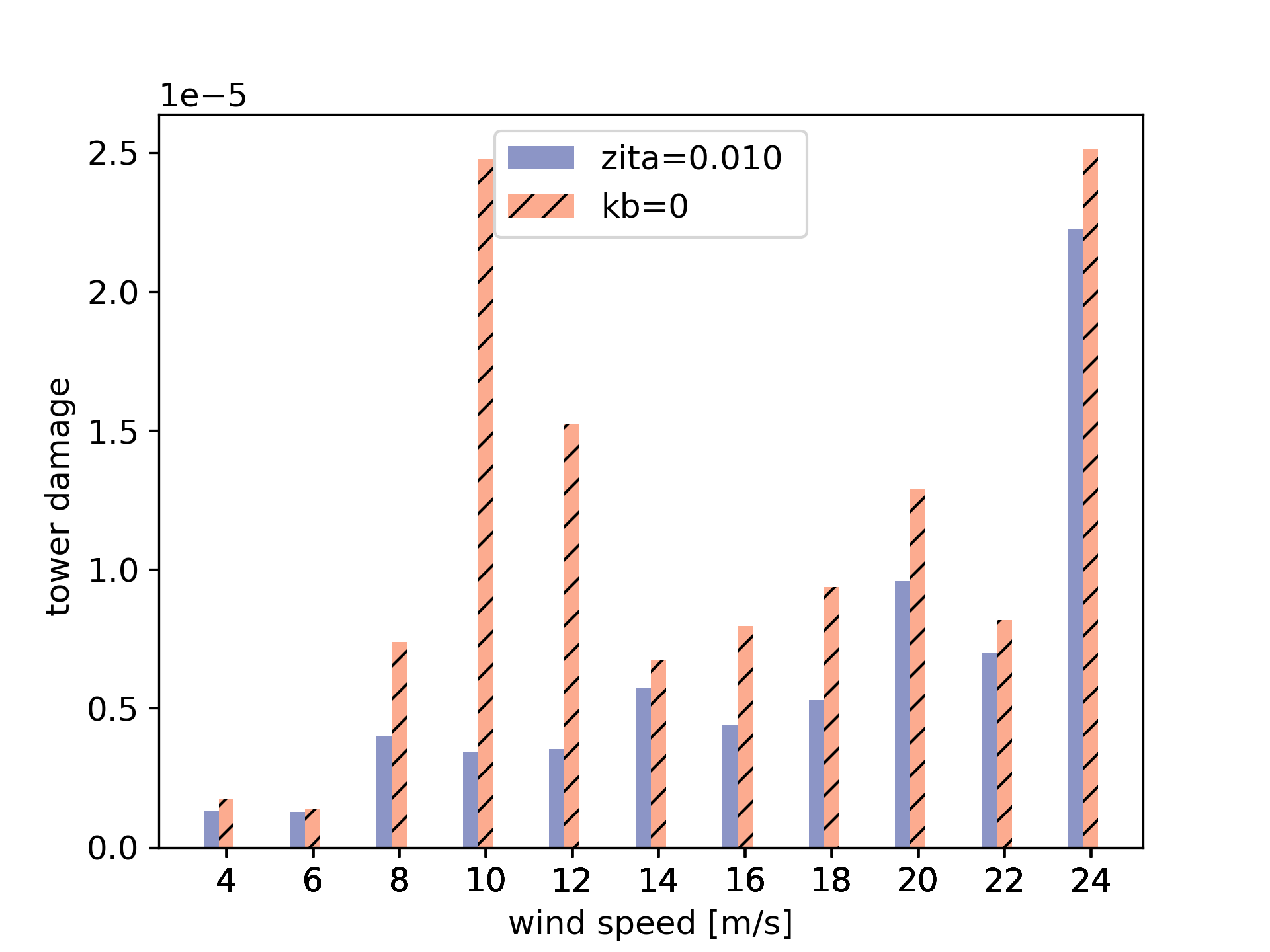}
\caption{Fatigue cumulative damage at tower bottom by using rainflow counting and linear Miner's rule. The damage is obtained considering the tower base design proposed by the UMaine, a Wohler's curve bi-linear with $m=3.0$ up to $10^6$ cycles and $m=5.0$ after, as proposed by DNV for Offshore steel. The probability of occurrence of each wind is equal, without any weibull distribution.}
\label{fig:damage}
\end{figure}

It is interesting to analyse Figure \ref{fig:damage}. 
This figure reports a deeper analysis of the fatigue damage. 
In fact, the stress in the tower bottom section is obtained by considering the design proposed by UMaine in \cite{UMaine}.
Then, an offshore Wohler's curve is considered with two slopes in the log-log domain: $m=3.0$ for loads with less than $1.0$ million cycles and $m=5.0$ for loads with higher number of cycles. 
Those are typical values proposed by DNV for offshore steel structures. 
This analysis leads to obtain an estimation of the $25$ years damage at the tower bottom. 
The gain is much more evident than the DEL. 
This is due to the second slope, $m=5.0$, which amplifies the changes in the load amplitudes. 
Offshore WTG in production are mostly subjected to a very high number of cycles of small amplitudes. 
This figure shows also the effect of the turbulence on the fatigue. 
In fact, looking at the reference, up to $12$ $ms^{-1}$, the shape of the damage distribution follows the one of the thrust curve. 
However, since the turbulence is a percentage of the average wind speed, from $16$ $ms^{-1}$, the damage starts increasing again.

\begin{table}[!h]
\label{tab:res_10ms}
\caption{Statistics for results concerning the case with mean wind speed at $10$ $ms^{-1}$. For each quantity of interest, there is the comparison of the minimum, maximum, mean and standard deviation values produced by the $\zeta_{plt}$-fixed control strategy and the reference.}
\centering
\begin{tabular}[t]{ p{2.4cm} | p{1.4cm}  p{1.4cm} | p{1.4cm}  p{1.4cm} | p{1.4cm}  p{1.4cm} | p{1.4cm}  p{1.4cm} }
\hline
       &  $\zeta_{plt}=0.1$  & $k_\beta=0$ &  $\zeta_{plt}=0.1$  & $k_\beta=0$ &  $\zeta_{plt}=0.1$  & $k_\beta=0$ &  $\zeta_{plt}=0.1$  & $k_\beta=0$ \\
      &  min & min & mean & mean & max & max & st.d. & st.d \\
\hline 
PtfmPitch [deg] & 2.84 & 1.09 & 4.55 & 4.56 & 5.62 & 5.91 & 0.51 & 0.59 \\
TwrBsM [MNm] & 165.40 & 80.97 & 366.66 & 367.79 & 493.80 & 505.10 & 43.20 & 46.82 \\
GenPwr [MW] & 7.85 & 5.49 & 12.09 & 12.01 & 15.15 & 15.19 & 1.29 & 1.47 \\
BldPitch [deg] & 0 & 0 & 0.218 & 0.176 & 6.651 & 8.33 & 0.394 & 0.729 \\
RotSpeed [rpm] & 5.23 & 5.34 & 6.90 & 6.92 & 7.90 & 8.00 & 0.61 & 0.60 \\
\end{tabular}
\label{tab:res_10ms}
\end{table}

\newpage

\conclusions  
The first part of this paper presents the analysis of the NMPZ related to the system of questions describing the dynamcs of a floating offshore wind turbine (FOWT). 
The equation of the rotor dynamics and the one of the platform dynamics are analysed in the complex domain to explicit the conditions leading to the two respective NMPZs. 
Two analytical examples are considered, one per each NMPZ condition, to compare the behaviour of a system affected by a NMPZ with another one outside the NMPZ domain. 
One of those NMPZ, instability given by the blade pitch on the rotor dynamics, is already known in literature and a compensation already exists to avoid it. 
Another one, related to instability given by the blade pitch to the platform dynamics, is a novelty in the community. The effects of the NMPZs are analysed on two analytical test cases: at the beginning both $\omega$ and $\dot \phi$ always converge to the right solutions just after the first steps. When both NMPZ conditions are not verified, those tendencies don't change. 
However, when the $\phi$-NMPZ is verified, $| \dot \omega |$ becomes so big that $\dot \phi$ jumps into unexpected values without converging to the expected solution. 
Similarly, when the $\omega$-NMPZ condition is verified, $| \dot \phi |$ becomes so big that $\omega$ oscillates before converging to the expected solution. 
NMPZs can cause important shifts and unexpected behaviors for both $\omega$ and $\phi$.

In the second part of the paper, the damping analysis is further investigated proposing a new strategy control for FOWT, named $ \zeta_{plt}$-fixed. 
This strategy focuses on the coupling between the rotor dynamics and the floating platform dynamics and it is based on a compensation parameter $k_{\beta}$ proportional to the platform pitch velocity. 
The idea behind the control strategy developed in this work is to use the blade pitch to damp the platform motions. 
An explicit expression linking $k_{\beta}$ to $ \zeta_{plt}$ (damping ratio imposed to the platform) is derived by defining a second order filter on the equation of the platform dynamics. 
This is different with respect to already existing strategies based on platform pitch compensation \cite{rosco, Stockhouse} which focus on decoupling rotor and platform dynamics. 
This difference is underlined by the values of $k_{\beta}$, which is negative for the new control strategy, while it is positive for the ones existing in literature. 
The performances of the $ \zeta_{plt}$-fixed strategy are tested analytically first and, then, numerically by considering an OpenFAST numerical twin of the Umaine IEA15MW FOWT. 
For a test representative of the DLC1.2, the $ \zeta_{plt}$-fixed strategy allows to reduce the loads at the tower foundation interface for all the considered wind speeds, without significant losses in terms of power production. 
The damage analysis shows a remarkable gain in terms of fatigue lifetime. 
The blade pitch use slightly increases remaining in the bounds of a standard controller limitations. 
For each FOWT system, some iterations are necessary in order to find the optimum value for $\zeta_{plt}$. 
This work highlights the importance of defining proper controller strategies for FOWT in order to reduce loads on the structure or improve the performances and, then, helps the industry to achieve the objective in terms of LCOE reduction. 









\noappendix       







\authorcontribution{Matteo Capaldo has contributed for the original idea of the new control strategy, the development of the numerical twin and the numerical tests and he is the main contributor for the paper editing. Paul Mella developed the mathematical framework, he has contributed for the original idea of the new control strategy and he has contributed for the paper editing} 

\competinginterests{The authors declare there are not competing interests in this work.} 

\end{document}